\documentclass[onecolumn]{svjour3}

\usepackage[dvipdfmx]{graphicx}
\usepackage{dcolumn}
\usepackage{bm}
\usepackage{amsmath}
\usepackage{amsfonts}
\usepackage{amssymb}
\usepackage{ascmac}
\usepackage{mathptmx}
\usepackage[letterpaper,mag=1400,includehead,margin=10truemm,tmargin=20truemm,lmargin=15truemm,rmargin=15truemm,twoside=false,truedimen]{geometry}

\journalname{Journal of Elasticity}

\begin{document}
\title{Revisiting stress propagation in a two-dimensional elastic circular disk under diametric loading}

\author{Yosuke Sato \and Haruto Ishikawa \and Satoshi Takada}

\institute{
Y. Sato \and H. Ishikawa
\at Department of Mechanical Systems Engineering, 
Tokyo University of Agriculture and Technology, 
2--24--16 Naka-cho, Koganei, Tokyo 184--8588, Japan
\and
S. Takada
\at Department of Mechanical Systems Engineering and Institute of Engineering, 
Tokyo University of Agriculture and Technology, 
2--24--16 Naka-cho, Koganei, Tokyo 184--8588, Japan,\\
\email{takada@go.tuat.ac.jp}, corresponding author
}

\date{Received: date / Accepted: date}

\maketitle

\begin{abstract}
In this paper, we present a comprehensive investigation of stress propagation in a two-dimensional elastic circular disk.
To accurately describe the displacements and stress fields within the disk, we employ a scalar and vector potential approach, representing them as sums of Bessel functions.
The determination of the coefficients for these expansions is accomplished in the Laplace space, where we compare the boundary conditions.
By converting the inverse Laplace transforms into complex integrals using residue calculus, we successfully derive explicit expressions for the displacements and stress fields.
Notably, these expressions encompass primary, secondary, and surface waves, providing a thorough characterization of the stress propagation phenomena within the disk.
Our findings contribute to the understanding of mechanical behavior in disk-shaped components and can be valuable in the design and optimization of such structures across various engineering disciplines.
\end{abstract}


\section{Introduction}
The comprehension of stress propagation within structural components stands as a paramount endeavor, as it not only facilitates the prognostication of mechanical behavior but also assures the secure and resourceful design of engineering systems \cite{Timoshenko,Fung,AkiRichards}.
In particular, the meticulous examination of stress distribution phenomena in a two-dimensional elastic disk under diametric loading has garnered considerable scholarly attention in recent research endeavors, owing to its pertinence across diverse engineering applications \cite{Guerrero19,Ramesh20,Ramesh22,Shins23}.

Prior investigations have delved into various facets of stress distribution analysis \cite{Ramesh20,Ramesh22,Shins23,Hung03,Schonert04,Ma08,Hua19}.
Primarily, researchers have scrutinized the determination of internal static stress distribution within a singular photoelastic material, employing principles rooted in elasticity theory \cite{Timoshenko,Fung,Landau,Frocht41,Coker57}.
Through the application of elasticity principles, it becomes conceivable to dissect stress distribution within a material, accounting for its mechanical attributes and external loading conditions \cite{Ramesh22,Shins23}.

Moreover, scholars have explored the nexus between stress distribution analysis and civil engineering, as well as other pertinent domains, such as concrete \cite{Wang14,Jiang20}.
Boussinesq's solution has been wielded to fathom stress distributions induced by applied loads on concrete structures \cite{Fung}.
Leveraging this solution, researchers gain valuable insights into the intricate stress states within concrete elements, thereby augmenting their grasp of structural comportment.

Furthermore, elasticity theory has manifestly emerged as a potent and versatile analytical tool in stress distribution analysis.
Its efficacy transcends static scenarios, enabling the analytical resolution of dynamic predicaments, including stress propagation within semi-infinite spaces.
Cagniard-de Hoop and Green's function methodologies have been harnessed to scrutinize stress propagation and other dynamic phenomena, capitalizing on the underpinnings of elasticity theory \cite{Fung,AkiRichards}.

However, in finite systems, attaining analytical solutions becomes progressively arduous.
The intricacies introduced by finite boundaries and varying material attributes present formidable hurdles \cite{Wu06}.
As a consequence, analytical solutions for stress distribution analysis in finite cases remain circumscribed, with only a select few specific systems having been successfully addressed \cite{Wu06,Jingu85,Jingu85_3D,Kessler91}.

While the Finite Element Method (FEM) analysis has emerged as a sanguine alternative, it is not devoid of limitations \cite{Hua15,Mazel16,Liu21}.
FEM analysis bestows the capacity to simulate stress distributions and behavior numerically across a spectrum of structures and materials.
Nevertheless, the pragmatic application of FEM analysis poses challenges, particularly in accurately capturing and dissecting the behavior of diverse wave types and dynamic phenomena.
Indeed, some papers \cite{Yu06,Kourkoulis12,Zhang20} indicate non-negligible discrepancies in stress distribution proximate to the point of load application, which comes from the fact that the number of meshes is insufficient.
This underscores the imperative for numerical methods to undergo rigorous benchmarking.
Furthermore, modern structures exhibit increasingly intricate geometries, rendering a plethora of benchmark tests highly desirable.

Hence, notwithstanding the mathematical exigencies entailed in theoretical analysis, a compelling impetus endures to advance the analytical comprehension of stress distribution.
Through the refinement of theoretical frameworks and methodologies, researchers can surmount the constraints of extant approaches, thereby attaining deeper insights into stress distributions within intricate systems.
This is vital as it enables a discourse on the underlying causes of observed behavior, which complements the perspectives offered by experiments and numerical computations.

In this exposition, we revisit the propagation of stress within a two-dimensional elastic circular disk.
It is noteworthy that the framework itself was adopted in a prior study \cite{Jingu85}.
However, a reexamination of this approach is warranted.
Foremost, their solutions harbor several discrepancies.
For instance, the static solution deviates from the classically established solution \cite{Timoshenko}.
Moreover, the time-evolving solutions are documented to diverge at the origin, a patently unphysical behavior.
Furthermore, the assorted wave types, such as P- and S-waves, constituting stress waves, have not been distinctly classified.
To rectify these issues, a reevaluation of this problem is imperative.

The organization of this manuscript unfolds as follows.
In the ensuing section, we provide a succinct elucidation of the model and setup under consideration.
Section~\ref{sec:elastodynamics} expounds on the derivation of displacement and stress within the framework of linearized elastodynamics, in the scenario where two forces act diametrically.
Subsequently, we expound upon the explicit formulations of the static and dynamic solutions in Sections~\ref{sec:static} and \ref{sec:dynamic}, respectively.
The acquired findings are elucidated and visually represented in Section~\ref{sec:results}.
In Section~\ref{sec:summary}, a synthesis of our results is presented, accompanied by a thorough discussion.
Appendix \ref{sec:derivation} provides concise insights into techniques employed in deriving the solution.
Appendix \ref{sec:Cartesian_polar} delineates the interrelationship between stress in Cartesian and polar coordinates.
Lastly, Appendix \ref{sec:Rayleigh} furnishes the derivation of the speed of the Rayleigh wave.

\section{Model and setup}
\begin{figure}[htbp]
    \centering
    \includegraphics[width=0.5\linewidth]{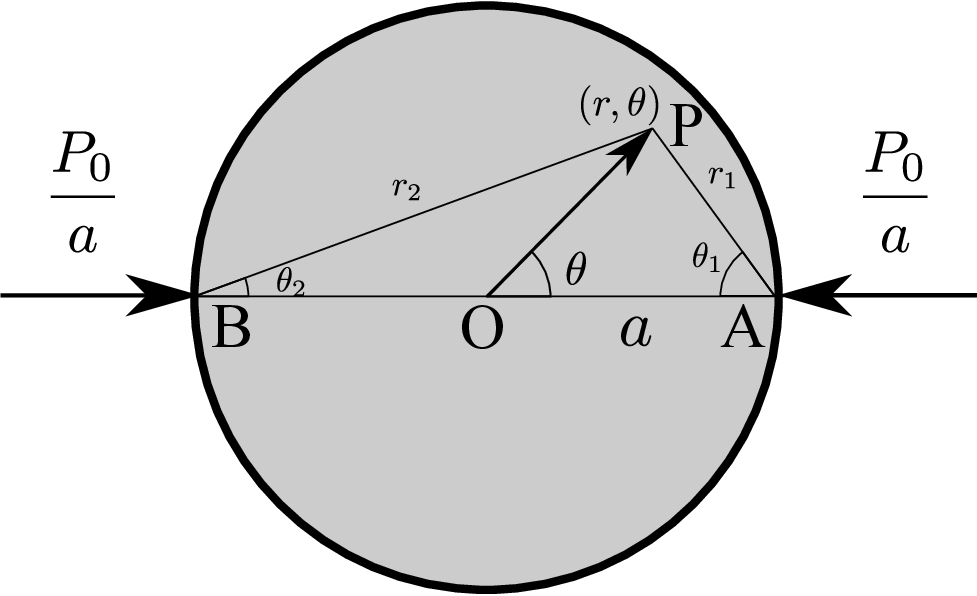}
    \caption{Schematic picture of our system.
    The stress acts on a circular disk whose diameter is $a$.}
    \label{fig:setup}
\end{figure}

Let us consider a two-dimensional circular disk whose radius is given by $a$ as shown in Fig.~\ref{fig:setup}.
We also assume that the mass density, shear modulus, and Poisson's ratio of the disk are, respectively, given by $\rho$, $G$, and $\nu$.
Here, we fix $\nu=0.3$ throughout this paper.
In this paper, we try to solve the time evolution of the stress when two identical but diametric loads act on the outer edge of the disk at $t=0$.
Without loss of generality, we put the magnitude of the loads as $P_0$ and their angles as $\theta=0$ and $\pi$.
Then, the boundary conditions of the stress components are written by
\begin{subequations}\label{eq:boundary_condition}
\begin{align}
    \left.\sigma_{rr}\right|_{r=a}
    &=-P(\theta)\Theta(t),\\
    \left.\sigma_{r\theta}\right|_{r=a}
    &=0,
\end{align}
\end{subequations}
with
\begin{align}
    P(\theta)
    &= \frac{P_0}{a}\left[\delta(\theta)+\delta(\theta-\pi)\right]\nonumber\\
    &= \frac{2P_0}{\pi a}\left[\frac{1}{2}+\sum_{m=2,4,\cdots}\cos(m\theta)\right],
    \label{eq:BC_diametric_P}
\end{align}
where $\delta(x)$ is the delta function.
In the next section, we solve the linearized elastodynamic equation under this boundary condition.

\section{Equations in linearized elastodynamics}\label{sec:elastodynamics}
In this section, we derive the equations of the displacement and stress in the framework of the linearized elastodynamics.
Now, we consider a situation under plane stress.
Under this situation, we solve the following Navier-Cauchy equation \cite{Fung}:
\begin{equation}
    \rho\frac{\partial^2}{\partial t^2}\bm{u}
    = G\nabla^2 \bm{u} + G\frac{1+\nu}{1-\nu}\bm\nabla \left(\bm\nabla \cdot \bm{u}\right).
    \label{eq:Navier-Cauchy}
\end{equation}

From the Helmholtz's theorem, let us write the displacement vector $u$ in terms of scalar and vector potentials ($\phi$ and $\bm{A}=(0, 0, A)^T$) as
\begin{equation}
    \bm{u}=\bm\nabla \phi + \bm\nabla \times \bm{A}.
    \label{eq:u_Helmholtz}
\end{equation}
We substitute Eq.~\eqref{eq:u_Helmholtz} into Eq.~\eqref{eq:Navier-Cauchy}, then we can rewrite Eq.~\eqref{eq:Navier-Cauchy} in terms of $\phi$ and $A$.
It is well known that Eq.~\eqref{eq:Navier-Cauchy} are satisfied when the scalar and the vector potential satisfy
\begin{equation}
    \nabla^2 \phi = \frac{1}{v_\mathrm{L}^2}\frac{\partial^2\phi}{\partial t^2},\quad
    \nabla^2 A = \frac{1}{v_{\mathrm T}^2}\frac{\partial^2A}{\partial t^2},
    \label{eq:phi_A_wave_eq}
\end{equation}
respectively \cite{Fung}, where we have introduced the sound speeds of the longitudinal ($v_\mathrm{L}$) and the transverse wave ($v_\mathrm{T}$) as
\begin{equation}
    v_\mathrm{L}\equiv \sqrt{\frac{2}{1-\nu}\frac{G}{\rho}},\quad
    v_\mathrm{T}\equiv
    \sqrt{\frac{G}{\rho}}.
\end{equation}
Using $\phi$ and $A$, each component of the strain and the stress in the polar coordinates is written as
\begin{subequations}\label{eq:u_sigma_phi_A}
\begin{align}
    u_r 
    &= \frac{\partial \phi}{\partial r}
    + \frac{1}{r}\frac{\partial A}{\partial \theta},\\
    u_\theta 
    &= \frac{1}{r}\frac{\partial \phi}{\partial \theta}
    - \frac{\partial A}{\partial r},\\
    \frac{\sigma_{rr}}{2G}
    &= - \frac{1}{r}\frac{\partial\phi}{\partial r}
    - \frac{1}{r^2}\frac{\partial^2\phi}{\partial\theta^2}
    + \frac{\partial}{\partial r}\left(\frac{1}{r}\frac{\partial A}{\partial \theta}\right)
    + \frac{1}{1-\nu}\nabla^2\phi,\\
    \frac{\sigma_{r\theta}}{2G}
    &= - \frac{\partial^2A}{\partial r^2}
    +  \frac{\partial}{\partial r}\left(\frac{1}{r}\frac{\partial \phi}{\partial \theta}\right)
    + \frac{1}{2}\nabla^2 A,\\
    \frac{\sigma_{\theta\theta}}{2G}
    &= \frac{1}{r}\frac{\partial \phi}{\partial r}
    + \frac{1}{r^2}\frac{\partial^2\phi}{\partial\theta^2}
    - \frac{\partial}{\partial r}
    \left(\frac{1}{r}\frac{\partial A}{\partial \theta}\right)
    + \frac{\nu}{1-\nu}\nabla^2\phi,
\end{align}
\end{subequations}

Now, it is convenient to consider the Laplace transforms of $\phi$ and $A$ as
\begin{equation}
    \overline{\phi}(s)\equiv \int_0^\infty \phi e^{-st}\mathrm{d}t, \quad
    \overline{A}(s)\equiv \int_0^\infty A e^{-st}\mathrm{d}t,
\end{equation}
where we regard $\overline{f}$ as the Laplace transform of $f$.
Assuming that the system remains still for $t<0$, Eqs.~\eqref{eq:phi_A_wave_eq} become
\begin{equation}
    \nabla^2 \overline{\phi} = \frac{s^2}{v_{\mathrm L}^2}\overline{\phi},\quad
    \nabla^2 \overline{A} = \frac{s^2}{v_{\mathrm T}^2}\overline{A},
    \label{eq:phi_A_wave_eq2}
\end{equation}
in the Laplace space.
Equations \eqref{eq:phi_A_wave_eq2} can be solved by separation of variables as
\begin{subequations}\label{eq:phi_bar_A_bar}
\begin{align}
    \overline{\phi}(s)
    &= \sum_{m=0}^\infty  a_m(s) I_m \left(\frac{r}{v_\mathrm{L}}s\right) \cos(m\theta),\\
    \overline{A}(s)
    &= \sum_{m=1}^\infty  b_m(s)I_m \left(\frac{r}{v_\mathrm{T}}s\right) \sin(m\theta),
\end{align}
\end{subequations}
where the coefficients $a_m(s)$ and $b_m(s)$ should be determined to satisfy the boundary conditions \eqref{eq:boundary_condition}, and $I_m(r)$ is the modified Bessel function of the first kind defined by \cite{Abramowitz}
\begin{equation}
    I_m (r)
    =\sum_{n=0}^\infty  \frac{1}{n! \Gamma(n+m+1)}\left(\frac{r}{2}\right)^{2n+m}.
\end{equation}
We note that we do not consider solutions that diverge at the origin, i.e., the modified Bessel function of the second kind $K_m(r)$ because the system should be kept finite in the whole range.

To determine $a_m$ and $b_m$, we first write the Laplace transforms of the displacement and the stress in the polar coordinates.
Substituting Eqs.~\eqref{eq:phi_bar_A_bar} into Eqs.~\eqref{eq:u_sigma_phi_A}, we get
\begin{subequations}\label{eq:usigma__Laplace}
\begin{align}
    ar^*\overline{u}_{r}
    &= \sum_{m=0}^\infty 
    \left[-F_{m,0}(r^*z)a_m(s) + m I_m (\mu r^*z)b_m(s)\right] \cos(m\theta),\\
    ar^*\overline{u}_{\theta}
    &= \sum_{m=1}^\infty
    \left[ -m I_m(r^*z)a_m(s)+F_{m,0}(\mu r^*z) b_m(s) \right] \sin(m\theta),\\
    \frac{a^2}{2G}r^{*2}\overline{\sigma}_{rr}
    &= \sum_{m=0}^\infty
    \left[F_{m,1}(r^*z)a_m(s) - F_{m,2}(\mu r^* z)b_m(s)\right] \cos(m\theta),\\
    \frac{a^2}{2G}r^{*2}\overline{\sigma}_{r\theta}
    &= \sum_{m=1}^\infty
    \left[F_{m,2}(r^* z) a_m(s)- F_{m,3}(\mu r^*z)b_m(s)\right] \sin(m\theta),\\
    \frac{a^2}{2G}r^{*2}\overline{\sigma}_{\theta\theta}
    &= \sum_{m=0}^\infty
    \left[F_{m,4}(r^*z) a_m(s) + F_{m,2}(\mu r^*z)b_m(s)\right] \cos(m\theta),
\end{align}
\end{subequations}
respectively, where we have introduced
\begin{subequations}
\begin{align}
    F_{m,0}(z)&\equiv -m I_m(z) - zI_{m+1}(z),\\
    F_{m,1}(z) 
    &\equiv
    \left[m(m-1)+\frac{z^2}{1-\nu}\right]I_m(z) - z I_{m+1}(z),\\
    F_{m,2}(z) 
    &\equiv -m(m-1)I_m(z) - m z I_{m+1}(z),\\
    F_{m,3}(z) 
    &\equiv \left[m(m-1)+\frac{z^2}{2}\right]I_m(z) - z I_{m+1}(z),\\
    F_{m,4}(z) 
    &\equiv \left[-m(m-1)+\frac{\nu z^2}{1-\nu}\right]I_m(z) + z I_{m+1}(z),
\end{align}
\end{subequations}
and the following dimensionless quantities for simplicity:
\begin{equation}
    r^*\equiv \frac{r}{a},\quad
    z \equiv \frac{sa}{v_{\rm L}},\quad
    t^* \equiv \frac{v_{\rm L}t}{a},\quad
    \mu\equiv \frac{v_\mathrm{L}}{v_\mathrm{T}}
    =\sqrt{\frac{2}{1-\nu}}(>1).
\end{equation}

From the boundary conditions \eqref{eq:boundary_condition}, the coefficients $a_m(s)$ and $b_m(s)$ become
\begin{subequations}\label{eq:a_b_0_diametric}
\begin{align}
    a_m(s)
    &=
    \begin{cases}
    \displaystyle -\frac{P_0 a^2}{2\pi G v_\mathrm{L}}\frac{1}{z F_{0,1}(z)} & (m=0)\\
    \displaystyle -\frac{P_0a^2}{\pi Gv_\mathrm{L}}
    \frac{F_{m,3}(\mu z)}{z D_m(z)} & (m=2,4,\cdots)\\
    0 & (m=1,3,\cdots)
    \end{cases},\\
    b_m(s) 
    &=
    \begin{cases}
    0 & (m=1, 3,\dots)\\
    \displaystyle -\frac{P_0a^2}{\pi Gv_\mathrm{L}}\frac{F_{m,2}(z)}{z D_m(z)} & (m=2,4,\cdots)
    \end{cases},
\end{align}
\end{subequations}
respectively, where we have introduced
\begin{equation}
    D_m(z) \equiv F_{m,1}(z) F_{m,3}(\mu z) - F_{m,2}(z) F_{m,2}(\mu z).
    \label{eq:def_D_m}
\end{equation}
Using these coefficients, the displacement and the stress become
\begin{equation}
    \begin{Bmatrix}
        \widetilde{u}_r \\
        \widetilde{\sigma}_{rr} \\ \widetilde{\sigma}_{\theta\theta}
    \end{Bmatrix}
    = \sum_{m=0,2,4,\cdots}^\infty
    \begin{Bmatrix}
        \widetilde{u}_r^{(m)} \\
        \widetilde{\sigma}_{rr}^{(m)} \\ \widetilde{\sigma}_{\theta\theta}^{(m)}
    \end{Bmatrix}
    \cos(m\theta),\quad
    \begin{Bmatrix}
        \widetilde{u}_\theta \\
        \widetilde{\sigma}_{r\theta}
    \end{Bmatrix}
    = \sum_{m=2,4,\cdots}^\infty
    \begin{Bmatrix}
        \widetilde{u}_\theta^{(m)} \\
        \widetilde{\sigma}_{r\theta}^{(m)}
    \end{Bmatrix}
    \sin(m\theta),
    \label{eq:summation}
\end{equation}
where we have introduced the scaled displacement and stress as
\begin{equation}
    \widetilde{u}_\alpha \equiv \frac{\pi G}{P_0}u_\alpha,\quad
    \widetilde{\sigma}_{\alpha\beta} \equiv \frac{\pi a}{P_0}\sigma_{\alpha\beta},
\end{equation}
with 
\begin{subequations}\label{eq:u_sigma_0}
\begin{align}
    \widetilde{u}_r^{(0)}
    &= \frac{1}{2r^*}\frac{1}{2\pi i}\int_\mathrm{Br} \frac{F_{0,0}(r^*z)}{z F_{0,1}(z)} \mathrm{e}^{t^*z}\mathrm{d}z,
    \label{eq:u_r_0}\\
    \widetilde{\sigma}_{rr}^{(0)}
    &= -\frac{1}{r^{*2}}\frac{1}{2\pi i}\int_\mathrm{Br} \frac{F_{0,1}(r^*z)}{z F_{0,1}(z)} \mathrm{e}^{t^*z}\mathrm{d}z,\\
    \widetilde{\sigma}_{\theta\theta}^{(0)}
    &= -\frac{1}{r^{*2}}\frac{1}{2\pi i}\int_\mathrm{Br} \frac{F_{0,4}(r^*z)}{z F_{0,1}(z)} \mathrm{e}^{t^*z}\mathrm{d}z,
\end{align}
\end{subequations}
and
\begin{subequations}\label{eq:u_sigma_m}
\begin{align}
    \widetilde{u}_r^{(m)} 
    &= \frac{1}{r^*}\frac{1}{2\pi i}\int_\mathrm{Br} 
    \frac{F_{m,0}(r^*z)F_{m,3}(\mu z)- m I_m(\mu r^*z)F_{m,2}(z)}{zD_m(z)}\mathrm{e}^{t^*z}\mathrm{d}z,
    \label{eq:u_r_m}\\
    \widetilde{u}_\theta^{(m)}
    &= \frac{1}{r^*} \frac{1}{2\pi i}\int_\mathrm{Br} 
    \frac{m I_m(r^* z)F_{m,3}(\mu z)- F_{m,0}(\mu r^*z)F_{m,2}(z)}{zD_m(z)}\mathrm{e}^{t^*z}\mathrm{d}z,\\
    \widetilde{\sigma}_{rr}^{(m)}
    &= -\frac{2}{r^{*2}}
    \frac{1}{2\pi i}\int_\mathrm{Br}
    \frac{F_{m,1}(r^*z)F_{m,3}(\mu z)- F_{m,2}(\mu r^*z)F_{m,2}(z)}{zD_m(z)}\mathrm{e}^{t^*z}\mathrm{d}z,
    \label{eq:sigma_rr_m}\\
    \widetilde{\sigma}_{r\theta}^{(m)}
    &= -\frac{2}{r^{*2}}
    \frac{1}{2\pi i}\int_\mathrm{Br}
    \frac{F_{m,2}(r^*z)F_{m,3}(\mu z)- F_{m,3}(\mu r^*z)F_{m,2}(z)}{zD_m(z)}\mathrm{e}^{t^*z}\mathrm{d}z,\\
    \widetilde{\sigma}_{\theta\theta}^{(m)}
    &= -\frac{2}{r^{*2}}
    \frac{1}{2\pi i}\int_\mathrm{Br}
    \frac{F_{m,4}(r^*z)F_{m,3}(\mu z)+F_{m,2}(\mu r^*z)F_{m,2}(z)}{zD_m(z)}\mathrm{e}^{t^*z}\mathrm{d}z,
\end{align}
\end{subequations}
for $m= 2, 4, \cdots$, respectively.
Here, $\int_\mathrm{Br}=\int_{\gamma-i\infty}^{\gamma+i\infty}$ is known as the Bromwich integral to calculate the inverse Laplace transforms, where $\gamma$ ($>0$) should be larger than the real part of any pole in the integrands of Eqs.~\eqref{eq:u_sigma_0} and \eqref{eq:u_sigma_m}.
These calculations can be done by using the fast Fourier transform method, while its numerical cost is not small \cite{NumericalRecipes}.
In the next section, on the other hand, we show a different method to evaluate these quantities by using the residue theorem \cite{Jingu85,Jingu85_3D}.
We note that we consider the scaled displacement $\widetilde{u}_\alpha$ and stress $\widetilde{\sigma}_{\alpha\beta}$ from now on.

\section{Static solution under diametric loads}\label{sec:static}
In this section, we solve a static problem.
Because we are interested in the steady states, that is, the long time limits of the above quantities, it is useful to use the final value theorem of the Laplace transform, i.e., $\lim_{t\to\infty}f(t)=\lim_{s\to0}[s \overline{f}(s)]$.
After some calculations, the final values are, respectively, given by
\begin{subequations}\label{eq:u_sigma_m_lim}
\begin{align}
    \lim_{t\to\infty}\widetilde{u}_r^{(0)} 
    &= -\frac{1-\nu}{1+\nu}\frac{r^*}{2},\quad
    \lim_{t\to\infty}\widetilde{\sigma}_{rr}^{(0)} 
    = \lim_{t\to\infty}\widetilde{\sigma}_{\theta\theta}^{(0)} = -1,
    \label{eq:u_sigma_st_0}\\
    \lim_{t\to\infty}
    \begin{Bmatrix}
        \widetilde{u}_r^{(m)} \\ \widetilde{u}_\theta^{(m)} 
    \end{Bmatrix}
    &= 
    \begin{Bmatrix}
        - \\ + 
    \end{Bmatrix}
    \frac{m}{2} \left(\frac{r^{*m-1}}{m-1}-\frac{r^{*m+1}}{m+1}\right)
    -
    \begin{Bmatrix}
        \displaystyle \frac{1-\nu}{1+\nu} \\ \displaystyle \frac{2}{1+\nu} 
    \end{Bmatrix}
    \frac{r^{*m+1}}{m+1}\quad (m=2,4,\cdots),\\
    \lim_{t\to\infty}
    \begin{Bmatrix}
        \widetilde{\sigma}_{rr}^{(m)} \\ \widetilde{\sigma}_{r\theta}^{(m)}\\
        \widetilde{\sigma}_{\theta\theta}^{(m)} 
    \end{Bmatrix}
    &= 
    \begin{Bmatrix}
        - \\ + \\ + 
    \end{Bmatrix}
    mr^{*m-2}(1-r^{*2}) 
    -2r^{*m}
    \begin{Bmatrix}
        1 \\ 0 \\ 1 
    \end{Bmatrix}\quad (m=2,4,\cdots),
\end{align}
\end{subequations}
(the detailed derivations are given in Appenix \ref{sec:derivation}).
Here, we have used the relation $\mu^2-1=(1+\nu)/(1-\nu)$.

\begin{figure}[htbp]
    \centering
    \includegraphics[width=\linewidth]{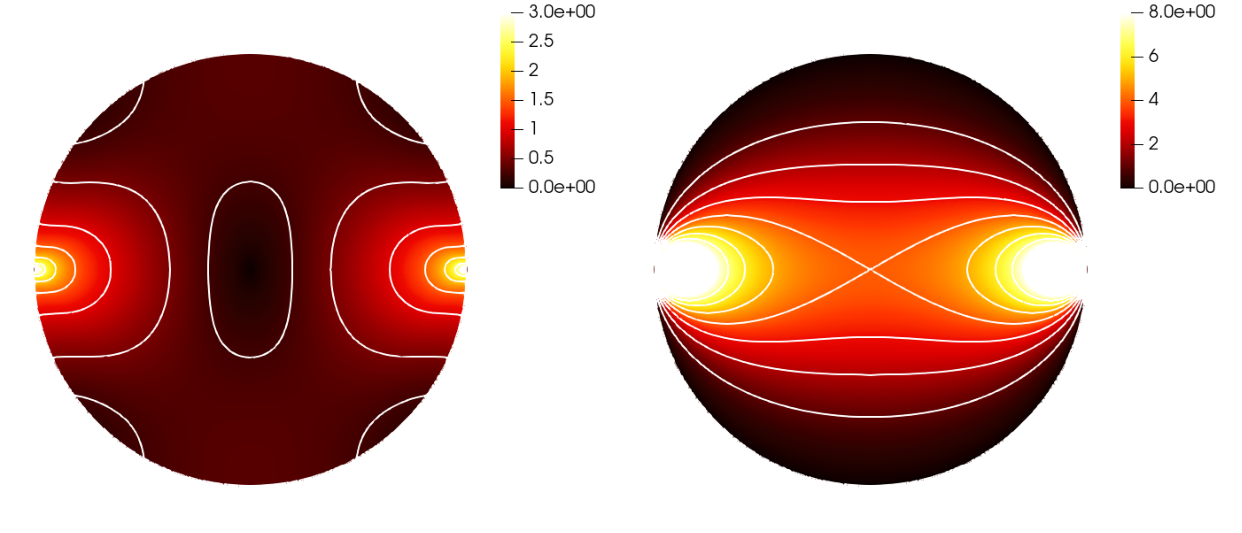}
    \caption{Contour plots of the scaled displacement $\widetilde{u}\equiv (\pi G/P_0)u$ and (scaled) principal stress difference $\widetilde{\sigma}_{1}-\widetilde{\sigma}_{2}\equiv (\pi G/P_0)(\sigma_{1}-\sigma_{2})$ for $\nu=0.3$.}
    \label{fig:static_u_sigma}
\end{figure}

Summing up each of Eqs.~\eqref{eq:u_sigma_m_lim} over $m$, we can obtain
\begin{subequations}\label{eq:u_sigma_st}
\begin{align}
    \widetilde{u}_r^{(\mathrm{st})}
    &\equiv
    \lim_{t\to\infty}\sum_{m=0,2,4,\cdots} \widetilde{u}_r^{(m)}\cos(m\theta)
    =
    \frac{1}{2}\frac{1-\nu}{1+\nu}r^*
    +\frac{1}{2}\sin\theta_1 \sin(\theta+\theta_1)
    +\frac{1}{2}\sin\theta_2 \sin(\theta-\theta_2)\nonumber\\
    &\hspace{13em}
    - \frac{1}{1+\nu}\cos\theta \log\frac{r_2^*}{r_1^*}
    - \frac{1-\nu}{1+\nu} \frac{\theta_1+\theta_2}{2}\sin\theta,\\
    \widetilde{u}_\theta^{(\mathrm{st})}
    &\equiv
    \lim_{t\to\infty}\sum_{m=2,4,\cdots} \widetilde{u}_\theta^{(m)}\sin(m\theta)
    =
    \frac{1}{2}\sin\theta_1 \cos(\theta+\theta_1)
    +\frac{1}{2}\sin\theta_2 \cos(\theta-\theta_2)\nonumber\\
    &\hspace{13em}
    + \frac{1}{1+\nu}\sin\theta \log\frac{r_2^*}{r_1^*} 
    - \frac{1-\nu}{1+\nu}\frac{\theta_1+\theta_2}{2}\cos\theta,\\
    \widetilde{\sigma}_{rr}^{(\mathrm{st})}
    &\equiv
    \lim_{t\to\infty}\sum_{m=0,2,4,\cdots} \widetilde{\sigma}_{rr}^{(m)}\cos(m\theta)
    =
    1-\frac{2\cos\theta_1 \cos^2(\theta+\theta_1)}{r_1^*}
    -\frac{2\cos\theta_2 \cos^2(\theta-\theta_2)}{r_2^*},
    \label{eq:sigma_rr_st}\\
    \widetilde{\sigma}_{r\theta}^{(\mathrm{st})}
    &\equiv
    \lim_{t\to\infty}\sum_{m=2,4,\cdots} \widetilde{\sigma}_{r\theta}^{(m)}\sin(m\theta)
    =
    \frac{\cos\theta_1 \sin[2(\theta+\theta_1)]}{r_1^*} +\frac{\cos\theta_2 \sin[2(\theta-\theta_2)]}{r_2^*},\\
    \widetilde{\sigma}_{\theta\theta}^{(\mathrm{st})}
    &\equiv
    \lim_{t\to\infty}\sum_{m=0,2,4,\cdots} \widetilde{\sigma}_{\theta\theta}^{(m)}\cos(m\theta)
    =
    1- \frac{2\cos\theta_1 \sin^2(\theta+\theta_1)}{r_1^*}
    - \frac{2\cos\theta_2 \sin^2(\theta-\theta_2)}{r_2^*},
\end{align}
\end{subequations}
respectively (see the detailed derivation in Appendix \ref{sec:derivation}), where we have introduced
\begin{subequations}\label{sec:def_r1_r2_theta1_theta2}
\begin{align}
    r_1^*&\equiv \frac{r_1}{a} 
    = \sqrt{1+r^{*2}-2r^* \cos\theta},\quad
    r_2^*\equiv \frac{r_2}{a}
    = \sqrt{1+r^{*2}+2r^* \cos\theta},\\
    \theta_1 &\equiv \tan^{-1}\frac{r^*\sin\theta}{1-r^*\cos\theta},\quad
    \theta_2 \equiv \tan^{-1}\frac{r^*\sin\theta}{1+r^*\cos\theta},
\end{align}
\end{subequations}
(see Fig.~\ref{fig:setup}).
Here, the superscript (st) is attached to indicate a steady solution.
These expressions are equivalent to those derived by the geometrical discussions \cite{Timoshenko}.
In this sense, we have successfully reformulate the results from the long time limit of the elastodynamic equation.
Figure \ref{fig:static_u_sigma} shows the evolution of the magnitude of the (scaled) displacement $\widetilde{u}$ and the (scaled) principal stress difference $\widetilde{\sigma}_{1}-\widetilde{\sigma}_{2}$ as \cite{Timoshenko}
\begin{align}
    \widetilde{u}&\equiv \sqrt{\widetilde{u}_r^2+\widetilde{u}_\theta^2},\quad
    \widetilde{\sigma}_{1}-\widetilde{\sigma}_{2} 
    = \sqrt{(\widetilde{\sigma}_{rr}-\widetilde{\sigma}_{\theta\theta})^2 
        + 4\widetilde{\sigma}_{r\theta}^2}.
    \label{eq:def_u_sigma}
\end{align}

We can also the expressions of the stress components in the Cartesian coordinates from those in the polar coordinates as discussed in Appendix~\ref{sec:Cartesian_polar}.
This shows that the extensional stress ($\widetilde{\sigma}_{yy}$) is constant along the line parallel to the loadings.
We note that this fact is the fundamental basis of the Brazillian test \cite{Goodman91,Vutukuri74}.

\section{Dynamic solution under diametric loads}\label{sec:dynamic}
In this section, we solve a dynamic problem.
In the following, we focus on $u_r$ to make the discussion simpler.
To calculate the above integrals, it is useful to use the residue theorem by closing the contour.
Now, the adding contour should be in the region of convergence.

\begin{figure}[htbp]
    \centering
    \includegraphics[width=0.5\linewidth]{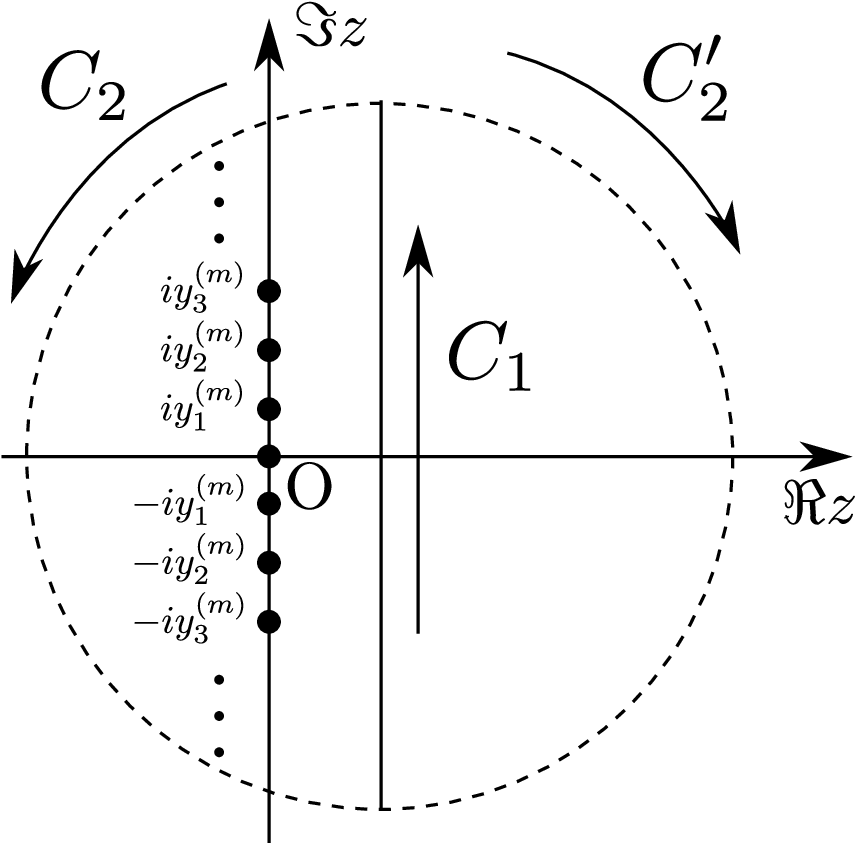}
    \caption{Integral paths for two cases: (i) $C\equiv C_1+C_2$ and (ii) $C^\prime \equiv C_1+C_2^\prime$ depending on the integrand.
    Points represent poles of the integrand.}
    \label{fig:path}
\end{figure}

Let us discuss which contours $C_2$ and $C_2^\prime$ in Fig.~\ref{fig:path} should be chosen depending on the time and the position.
For $|z|\to \infty$, the integrand of Eq.~\eqref{eq:u_r_0} and the first one of Eq.~\eqref{eq:u_r_m}, and  the second one of Eq.~\eqref{eq:u_r_m} behave as $e^{(r^*-1+t^*)z}/z$ and $e^{(\mu(r^*-1)+t^*)z}/z$, respectively.
Depending on the relationship between $r^*$ and $t^*$, the results changes as shown in the following subsections.

\subsection{Region I ($t^*<1-r^*$)}
First, let us consider the case for $t^*<1-r^*$.
In the following, we call this region as region I, and we put the subscript $\mathrm{I}$ to the quantities.
In this region, both terms converge to zero if we choose the right half-plane, which means that the Bromwich integral can be converted to the integral with the path $C_1+C_2$.
Because there are no pole of the integrand in this path, this integral becomes zero.
This corresponds to a situation where neither P- nor S-wave arrives at the position $r^*$ at time $t^*$.
Therefore, we get
\begin{equation}
    \widetilde{u}_{r,\mathrm{I}}^{(m)} = 0,
\end{equation}
for $m\ge 0$.

\subsection{Region II ($1-r^*\le t^*<\mu(1-r^*)$)}
\begin{figure}[htbp]
    \centering
    \includegraphics[width=0.6\linewidth]{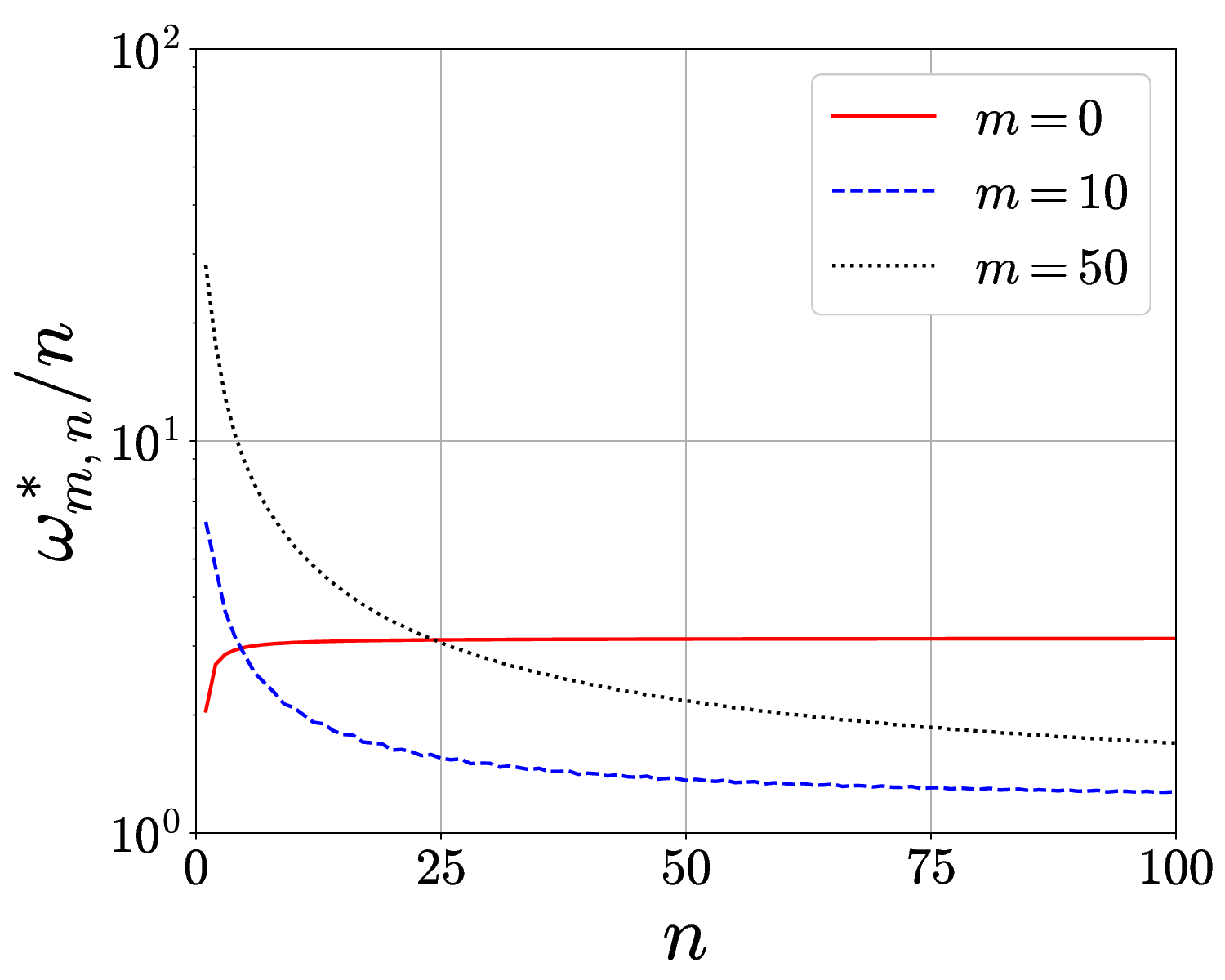}
    \caption{Plots of $\omega_{m,n}/n$ against $n$ for $m=0$, $10$, and $50$ when we fix $\nu=0.3$.}
    \label{fig:y_mn}
\end{figure}

Next, let us consider the case for $1-r^*\le t^*<\mu(1-r^*)$, which we call region II.
Because only the P-wave arrives in this region, we put the subscript $\mathrm{II}$ to the quantities.
The former and the latter converge to zero in the left ($C_2$ in Fig.~\ref{fig:path}) and the right half-plane ($C_2^\prime$), respectively.
In this case, only the former survives, which means that only the P-wave arrives at this position.
Then, we can obtain
\begin{equation}
    \widetilde{u}_{r,\mathrm{II}}^{(m)}
    =
    \begin{cases}
    \displaystyle \sum_{n=0}^\infty \mathrm{Res}_{z=z_n^{(0)}} \frac{F_{0,0}(r^*z)}{2r^*z F_{0,1}(z)} \mathrm{e}^{t^*z} & (m=0),\\
    \displaystyle \sum_{n=0}^\infty \mathrm{Res}_{z=z_n^{(m)}}  \frac{F_{m,0}(r^*z)F_{m,3}(\mu z)}{r^*zD_m(z)}\mathrm{e}^{t^*z} & (m=2,4,\cdots),
    \end{cases}
\end{equation}
where $\mathrm{Res}_{z=a}f(z)$ represents a residue at $z=a$.
As far as we have investigated, both $F_{0,1}(r^*, z)$ and $D_m(z)$ have only pure imaginary roots.
Figure \ref{fig:y_mn} shows the values of $\omega_{m,n}^*$ for $m=0$, $10$, and $50$ when we fix $\nu=0.3$.
Once we fix the Poisson's ratio $\nu$, we can numerically obtain a list of $\omega_{m,n}^*$ up to an arbitrary number of $n$ by using the Newton-Raphson method~\cite{NumericalRecipes}.
In addition, it is easy to show that $z=-i y$ is also a solution of either of them if $z=+iy$ is its solution.
Now, we put these roots as $\pm i\omega_{m,n}^*$ ($n=0, 1, 2, \cdots, \omega_{m,n}^*< \omega_{m,n+1}^*$), where we have regarded $\omega_{m,0}^*=0$.
Then, after some calculations, we get
\begin{equation}
    \widetilde{u}_{r,\mathrm{II}}^{(m)}
    =\widetilde{u}_{r,\mathrm{P}}^{(m,0)}(r^*,t^*)
    + \sum_{n=1}^\infty \widetilde{u}_{r,\mathrm{P}}^{(m)}(r^*,\omega_{m,n}^*)\cos(\omega_{m,n}^* t^*),
    \label{eq:u_r_II}
\end{equation}
with 
\begin{subequations}\label{eq:ur0_urm_residues}
\begin{align}
    \widetilde{u}_{r,\mathrm{P}}^{(m,0)}(r^*,t^*)
    &= 
    \begin{cases}
        \displaystyle \mathrm{Res}_{z=0}\frac{F_{0,0}(r^*z)}{2r^*z F_{0,1}(z)} \mathrm{e}^{t^*z} & (m=0),\\
        \displaystyle \mathrm{Res}_{z=0}\frac{F_{m,0}(r^*z)F_{m,3}(\mu z)}{r^*zD_m(z)}\mathrm{e}^{t^*z} & (m=2,4,\cdots),
    \end{cases}\\
    \widetilde{u}_{r,\mathrm{P}}^{(m)}(r^*,\omega^*)
    &=
    \begin{cases}
        \displaystyle \frac{f_{0,0}(r^* \omega^*)}{r^*\omega^* g_{0,1}(\omega^*)} & (m=0),\\
        \displaystyle \frac{2f_{m,0}(r^*\omega^*)f_{m,3}(\mu \omega^*)}{r^*\omega^*d_m(\omega^*)} & (m=2,4,\cdots),
    \end{cases}
\end{align}
\end{subequations}
where we have introduced
\begin{subequations}
\begin{align}
    f_{m,j}(y)&\equiv i^{-m}F_{m,j}(iy),\\
    g_{m,j}(y)&\equiv i^{-(m-1)}F_{m,j}^\prime (iy),\\
    d_m(y)&\equiv i^{-(2m-1)}D_m^\prime(iy),
\end{align}
\end{subequations}
($j=0, 1, \cdots, 5$), where their explicit expressions are presented in Table \ref{fig:func_list}.
After some calculations, one obtains
\begin{align}
    \widetilde{u}_r^{(\mathrm{tr})}
    &\equiv \sum_{m=0,2,4,\cdots} \widetilde{u}_{r,\mathrm{P}}^{(m,0)}\cos(m\theta)\nonumber\\
    &=\frac{1}{2}\frac{1-\nu}{1+\nu}r^* 
    + \frac{1-\nu}{1+\nu}\frac{1-r^{*2}}{2r^*r_1^*r_2^*}\cos(\theta_1-\theta_2)
    - \frac{1-\nu}{1+\nu}\frac{2r^*t^{*2}}{r_1^{*2}r_2^{*2}}\cos[2(\theta+\theta_1-\theta_2)]\nonumber\\
    &\hspace{1em}
    -\frac{2(3-\nu)+(1-\nu^2)r^{*2}}{4(1+\nu)^2}\frac{1}{r^{*2}}\left[\cos\theta \log\frac{r_2^*}{r_1^*}+(\theta_1+\theta_2)\sin\theta\right]\nonumber\\
    &\hspace{1em}
    -\frac{5-2\nu+\nu^2}{2(1+\nu)^2}\frac{1}{r^{*3}}\left[\cos(2\theta)\log(r_1^*r_2^*)-(\theta_1-\theta_2)\sin(2\theta) \right],
    \label{eq:u_r_tr}
\end{align}
where the derivation is similar to $\widetilde{u}_r^{(\mathrm{st})}$ given in Appendix \ref{sec:derivation}.


\subsection{Region III ($t^*\ge \mu(1-r^*)$)}
Third, let us consider the case for $t^*\ge \mu(1-r^*)$.
In this case, both the P- and S-waves arrives at the position $(r,\theta)$.
For simplicity, we put the subscript PS.
In this case, both terms converges to zero in the left half-plane, and can be evaluated by the residue theorem.
Interestingly, the poles obtained in the previous subsection are also poles in this region.
Using the similar procedure as that in the previous subsection, we find that $u_r^{(0)}$ is equivalent to Eq.~\eqref{eq:u_r_II} and $\widetilde{u}_r^{(m)}$ ($m\ge 1$) becomes
\begin{align}
    \widetilde{u}_{r,\mathrm{III}}^{(m,0)}
    &= \mathrm{Res}_{z=0}\frac{F_{m,0}(r^*z)F_{m,3}(\mu z)- m I_m(\mu r^*z)F_{m,2}(z)}{zD_m(z)}\mathrm{e}^{t^*z},\\
    \widetilde{u}_{r,\mathrm{III}}^{(m,n)}
    &= \widetilde{u}_{r,\mathrm{P}}^{(m,n)}
    - \frac{m J_m(\mu r^*y)f_{m,2}(y)}{\omega_{m,n}^* d_m(\omega_{m,n}^*)}.
\end{align}
Here, it is straightforward to obtain that $\sum_{m=0,2,4,\cdots}\widetilde{u}_{r,\mathrm{PS}}^{(m,0)}\cos(m\theta)$ is equivalent to the static solution obtained in the previous section.

Let us summarize our results:
\begin{subequations}\label{eq:u_sigma_final}
\begin{align}
    \widetilde{u}_r &=
    \Theta_\mathrm{P}(1-\Theta_\mathrm{S}) \widetilde{u}_r^{(\mathrm{tr})}(r^*,\theta,t^*)
    + \Theta_\mathrm{S} \widetilde{u}_r^{(\mathrm{st})}(r^*,\theta)\nonumber\\
    &\hspace{1em}
    + \sum_{m=0,2,4,\cdots} \sum_{n=1}^\infty
    \left[\Theta_\mathrm{P}\widetilde{u}_{r,\mathrm{P}}^{(m)}(r^*,\omega_{m,n}^*)
    +\Theta_\mathrm{S}\widetilde{u}_{r,\mathrm{S}}^{(m)}(r^*,\omega_{m,n}^*)\right]
    \cos(m\theta) \cos(\omega_{m,n}^*t^*),
    \label{eq:u_r}\\
    \widetilde{u}_\theta &=
    \Theta_\mathrm{P}(1-\Theta_\mathrm{S}) \widetilde{u}_\theta^{(\mathrm{tr})}(r^*,\theta,t^*)
    + \Theta_\mathrm{S} \widetilde{u}_\theta^{(\mathrm{st})}(r^*,\theta)\nonumber\\
    &\hspace{1em}
    + \sum_{m=2,4,\cdots} \sum_{n=1}^\infty
    \left[\Theta_\mathrm{P}\widetilde{u}_{\theta,\mathrm{P}}^{(m)}(r^*,\omega_{m,n}^*)
    +\Theta_\mathrm{S}\widetilde{u}_{\theta,\mathrm{S}}^{(m)}(r^*,\omega_{m,n}^*)\right] \sin(m\theta) \cos(\omega_{m,n}^*t^*),
    \label{eq:u_t}\\
    \widetilde{\sigma}_{rr} &=
    \Theta_\mathrm{P}(1-\Theta_\mathrm{S}) \widetilde{\sigma}_{rr}^{(\mathrm{tr})}(r^*,\theta,t^*)
    + \Theta_\mathrm{S} \widetilde{\sigma}_{rr}^{(\mathrm{st})}(r^*,\theta)\nonumber\\
    &\hspace{1em}
    + \sum_{m=0,2,4,\cdots} \sum_{n=1}^\infty
    \left[\Theta_\mathrm{P}\widetilde{\sigma}_{rr,\mathrm{P}}^{(m)}(r^*,\omega_{m,n}^*)
    + \Theta_\mathrm{S}\widetilde{\sigma}_{rr,\mathrm{S}}^{(m)}(r^*,\omega_{m,n}^*)\right] \cos(m\theta) \cos(\omega_{m,n}^*t^*),
    \label{eq:sigma_rr}\\
    \widetilde{\sigma}_{r\theta} &=
    \Theta_\mathrm{P}(1-\Theta_\mathrm{S}) \widetilde{\sigma}_{r\theta}^{(\mathrm{tr})}(r^*,\theta,t^*)
    + \Theta_\mathrm{S} \widetilde{\sigma}_{r\theta}^{(\mathrm{st})}(r^*,\theta)\nonumber\\
    &\hspace{1em}
    + \sum_{m=2,4,\cdots} \sum_{n=1}^\infty
    \left[\Theta_\mathrm{P}\widetilde{\sigma}_{r\theta,\mathrm{P}}^{(m)}(r^*,\omega_{m,n}^*)
    + \Theta_\mathrm{S}\widetilde{\sigma}_{r\theta,\mathrm{S}}^{(m)}(r^*,\omega_{m,n}^*)\right] \sin(m\theta) \cos(\omega_{m,n}^*t^*),
    \label{eq:sigma_rt}\\
    \widetilde{\sigma}_{\theta\theta} &=
    \Theta_\mathrm{P}(1-\Theta_\mathrm{S}) \widetilde{\sigma}_{\theta\theta}^{(\mathrm{tr})}(r^*,\theta,t^*)
    + \Theta_\mathrm{S} \widetilde{\sigma}_{\theta\theta}^{(\mathrm{st})}(r^*,\theta)\nonumber\\
    &\hspace{1em}
    + \sum_{m=0,2,4,\cdots} \sum_{n=1}^\infty
    \left[\Theta_\mathrm{P}\widetilde{\sigma}_{\theta\theta,\mathrm{P}}^{(m)}(r^*,\omega_{m,n}^*)
    + \Theta_\mathrm{S}\widetilde{\sigma}_{\theta\theta,\mathrm{S}}^{(m)}(r^*,\omega_{m,n}^*)\right] \cos(m\theta) \cos(\omega_{m,n}^*t^*),
    \label{eq:sigma_tt}
\end{align}
\end{subequations}
where $\Theta_\mathrm{P}$ and $\Theta_\mathrm{S}$ are, respectively, abbreviations of 
\begin{subequations}
\begin{align}
    \Theta_\mathrm{P}&\equiv \Theta\left(t^*-(1-r^*)\right),\\
    \Theta_\mathrm{S}&\equiv \Theta\left(t^*-\mu(1-r^*)\right).
\end{align}
\end{subequations}
Here, the detailed expressions are listed in Tables \ref{fig:u_expressions} and \ref{fig:sigma_expressions}.
We note that $\Theta_\mathrm{P}(1-\Theta_\mathrm{S})$ in Eqs.~\eqref{eq:u_sigma_final} stands for that this term only survives after the P-wave arrives but the S-wave does not.

\begin{figure}[htbp]
    \centering
    \includegraphics[width=\linewidth]{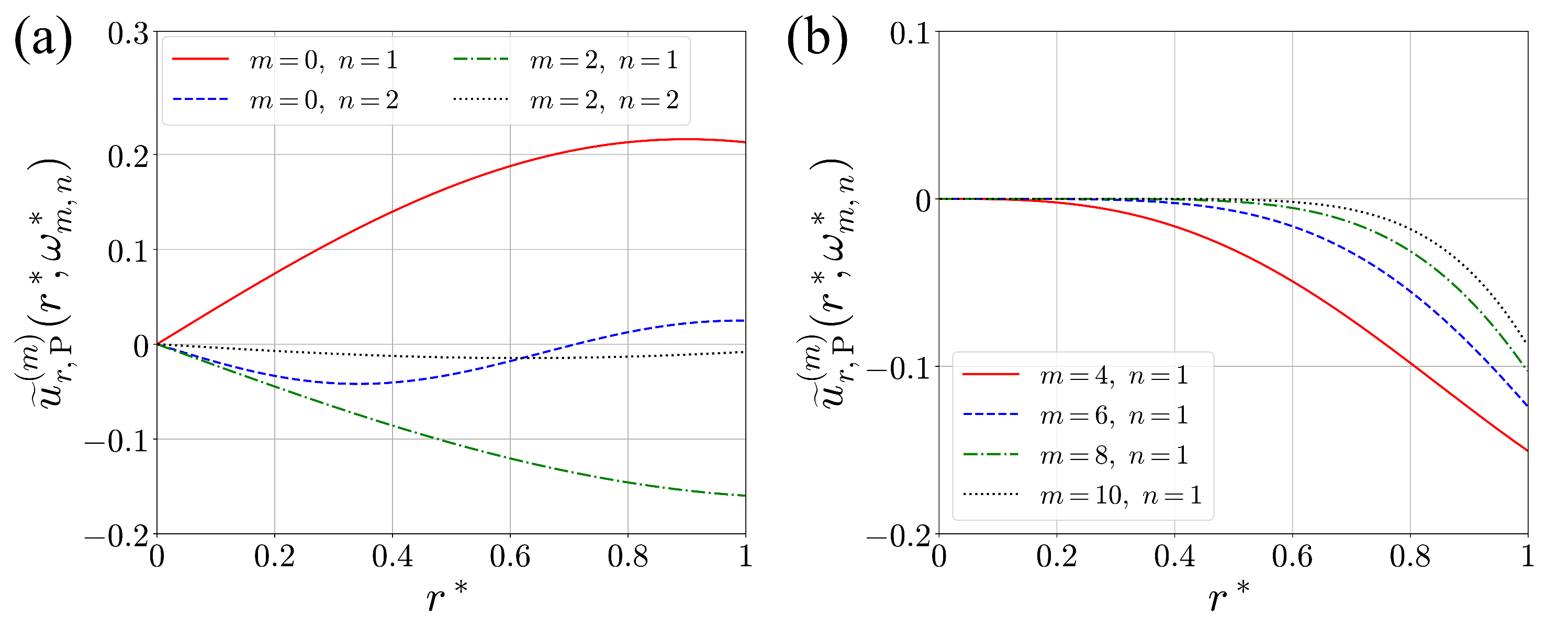}
    \caption{Plots of $\widetilde{u}_r^{(m)}(r^*,\omega_{m,n}^*)$ against $r^*$ for (a) various $n$ with fixed values of $m$ and (b) various $m$ with $n=1$.
    Here, we fix $\nu=0.3$.}
    \label{fig:u_sigma_mn_r}
\end{figure}

Figure \ref{fig:u_sigma_mn_r} shows the profiles of $\widetilde{u}_r^{(m)}(r^*,\omega_{m,n}^*)$ for various sets of $(m, n)$.
For larger $m$ and $n$, the values converge to zero, which means that the truncation of Eq.~\eqref{eq:u_sigma_final} up to $M_\mathrm{max}$ and $N_\mathrm{max}$ with respect to $m$ and $n$, respectively, also converge when we choose sufficient large values of $M_\mathrm{max}$ and $N_\mathrm{max}$.
In the actual calculations, we use $M_\mathrm{max}=N_\mathrm{max}=200$.

\begin{table}[htbp]
    \caption{Expressions of $f_{m,i}(y)$, $g_{m,i}(y)$, $n_{m,i}^{(\mathrm{P})}(y)$, $n_{m,i}^{(\mathrm{S})}(y)$ ($i=0$--$4$), and $d_m(y)$ that appear in the main text.}
    \label{fig:func_list}
    \centering
    \begin{tabular}{c|l}\hline\hline
    $f_{m,0}(y)$ & $\displaystyle -m J_m(y) + yJ_{m+1}(y)$\\ 
    $f_{m,1}(y)$ & $\displaystyle \left[m(m-1)-\frac{y^2}{1-\nu}\right]J_m(y) +yJ_{m+1}(y)$\\ 
    $f_{m,2}(y)$ & $\displaystyle -m(m-1)J_m(y) + m y J_{m+1}(y)$\\
    $f_{m,3}(y)$ & $\displaystyle \left[m(m-1)-\frac{y^2}{2}\right]J_m(y) + y J_{m+1}(y)$\\
    $f_{m,4}(y)$ & $\displaystyle \left[-m(m-1)-\frac{\nu y^2}{1-\nu}\right]J_m(y) - y J_{m+1}(y)$\\
    \hline
    $g_{m,0}(y)$ & $\displaystyle -\frac{m^2-y^2}{y}J_m(y)$\\
    $g_{m,1}(y)$ & $\displaystyle \frac{1}{y}\left[m^2(m-1) - \frac{m+1+\nu}{1-\nu}y^2\right] I_m(z) -\left(m^2 - \frac{y^2}{1-\nu}\right) J_{m+1}(y)$\\
    $g_{m,2}(y)$ & $\displaystyle -\frac{m}{y} [m(m-1)-y^2] J_m(y) - m J_{m+1}(y)$\\
    $g_{m,3}(y)$ & $\displaystyle \frac{m}{y}\left[m(m-1)-\frac{y^2}{2}\right] J_m(y)
    -\left(m^2-\frac{y^2}{2}\right) J_{m+1}(y)$\\
    $g_{m,4}(y)$ & $\displaystyle -\frac{1}{y}\left[m^2(m-1) + \frac{1+\nu+m\nu}{1-\nu}y^2\right] J_m(y) +\left(m^2+\frac{\nu}{1-\nu}y^2\right) J_{m+1}(y)$\\
    \hline
    $d_m(y)$ & $g_{m,1}(y)f_{m,3}(\mu y) 
    + \mu f_{m,1}(y) g_{m,3}(\mu y) - g_{m,2}(y) f_{m,2}(\mu y) 
    - \mu f_{m,2}(y) g_{m,2}(\mu y)$\\
    \hline
    \hline
    \end{tabular}
\end{table}

\begin{table}[htbp]
    \caption{Expressions of $\widetilde{u}_\alpha^{(\mathrm{tr})}$, $\widetilde{u}_\alpha^{(\mathrm{st})}$, $\widetilde{u}_{\alpha,\mathrm{P}}^{(m)}$, and $\widetilde{u}_{\alpha,\mathrm{S}}^{(m)}$ in Eqs.\ \eqref{eq:u_r} and \eqref{eq:u_t}.}
    \label{fig:u_expressions}
    \centering
    \begin{tabular}{c|l|c|l}\hline\hline
    $\widetilde{u}_r^{(\mathrm{tr})}$
    & \multicolumn{3}{l}{$\displaystyle \frac{1}{2}\frac{1-\nu}{1+\nu}r^* 
    + \frac{1-\nu}{1+\nu}\frac{1-r^{*2}}{2r^*r_1^*r_2^*}\cos(\theta_1-\theta_2)
    - \frac{1-\nu}{1+\nu}\frac{2r^*t^{*2}}{r_1^{*2}r_2^{*2}}\cos[2(\theta+\theta_1-\theta_2)]$}\\
    & \multicolumn{3}{l}{$\displaystyle -\frac{2(3-\nu)+(1-\nu^2)r^{*2}}{4(1+\nu)^2}\frac{1}{r^{*2}}\left[\cos\theta \log\frac{r_2^*}{r_1^*}+(\theta_1+\theta_2)\sin\theta\right]$}\\
    & \multicolumn{3}{l}{$\displaystyle -\frac{5-2\nu+\nu^2}{2(1+\nu)^2}\frac{1}{r^{*3}}\left[\cos(2\theta)\log(r_1^*r_2^*)-(\theta_1-\theta_2)\sin(2\theta) \right]$}\\ 
    \hline
    $\widetilde{u}_r^{(\mathrm{st})}$ 
    & \multicolumn{3}{l}{$\displaystyle \frac{1}{2}\frac{1-\nu}{1+\nu}r^*
    + \frac{1}{2}\sin\theta_1 \sin(\theta+\theta_1)
    + \frac{1}{2}\sin\theta_2 \sin(\theta-\theta_2)
    -\frac{1}{1+\nu}\cos\theta \log\frac{r_2^*}{r_1^*}
    - \frac{1}{2}\frac{1-\nu}{1+\nu} (\theta_1+\theta_2)\sin\theta$}\\
    \hline
    $\widetilde{u}_{r,\mathrm{P}}^{(m)}$ 
    & $\begin{cases}
        \displaystyle \frac{f_{0,0}(r^* \omega^*)}{r^*\omega^* g_{0,1}(\omega^*)} & (m=0)\\
        \displaystyle \frac{2f_{m,0}(r^*\omega^*)f_{m,3}(\mu \omega^*)}{r^*\omega^*d_m(\omega^*)} & (m\ge2)
    \end{cases}$
    & $\widetilde{u}_{r,\mathrm{S}}^{(m)}$ 
    & $\begin{cases}
        0 & (m=0)\\
        \displaystyle -\frac{2mJ_m(\mu r^* \omega^*) f_{m,2}(\omega^*)}{r^*\omega^*d_m(\omega^*)} & (m\ge2)
    \end{cases}$\\
    \hline\hline
    $\widetilde{u}_\theta^{(\mathrm{tr})}$ 
    & \multicolumn{3}{l}{$\displaystyle - \frac{1-\nu}{1+\nu}\frac{1-r^{*2}}{2r^*r_1^*r_2^*}\sin(\theta_1-\theta_2)
    + \frac{1-\nu}{1+\nu}\frac{2r^*t^{*2}}{r_1^{*2}r_2^{*2}}\sin[2(\theta+\theta_1-\theta_2)]$}\\
    &\multicolumn{3}{l}{$\displaystyle +\frac{2(3-\nu)+(1-\nu^2)r^{*2}}{4(1+\nu)^2}
    \frac{1}{r^{*2}}\left[-\sin\theta\log\frac{r_2^*}{r_1^*} + (\theta_1+\theta_2)\cos\theta\right]$}\\
    &\multicolumn{3}{l}{$\displaystyle -\frac{5-2\nu+\nu^2}{2(1+\nu)^2}
    \frac{1}{r^{*3}}\left[\sin(2\theta)\log(r_1^*r_2^*)+(\theta_1-\theta_2)\cos(2\theta)\right]$}\\
    \hline
    $\widetilde{u}_\theta^{(\mathrm{st})}$ 
    & \multicolumn{3}{l}{$\displaystyle \frac{1}{2}\sin\theta_1 \cos(\theta+\theta_1)
    + \frac{1}{2}\sin\theta_2 \cos(\theta-\theta_2)
    + \frac{1}{1+\nu}\sin\theta \log\frac{r_2^*}{r_1^*} 
    - \frac{1}{2}\frac{1-\nu}{1+\nu}(\theta_1+\theta_2)\cos\theta$}\\
    \hline
    $\widetilde{u}_{\theta,\mathrm{P}}^{(m)}$ 
    & $\displaystyle \frac{2mJ_m(r^* \omega^*)f_{m,3}(\mu\omega^*)}{r^*\omega^*d_m(\omega^*)}$
    & $\widetilde{u}_{\theta,\mathrm{S}}^{(m)}$ 
    & $\displaystyle -\frac{2f_{m,0}(\mu r^* \omega^*) f_{m,2}(\omega^*)}{r^*\omega^*d_m(\omega^*)}$\\
    \hline\hline
    \end{tabular}
\end{table}
\begin{table}[htbp]
    \caption{Expressions of $\widetilde{\sigma}_{\alpha\beta}^{(\mathrm{tr})}$, $\widetilde{\sigma}_{\alpha\beta}^{(\mathrm{st})}$, $\widetilde{\sigma}_{\alpha\beta,\mathrm{P}}^{(m)}$, and $\widetilde{\sigma}_{\alpha\beta,\mathrm{S}}^{(m)}$ in Eqs.\ \eqref{eq:sigma_rr}--\eqref{eq:sigma_tt}.}
    \label{fig:sigma_expressions}
    \centering
    \begin{tabular}{c|l|c|l}\hline\hline
    $\widetilde{\sigma}_{rr}^{(\mathrm{tr})}$ 
    & \multicolumn{3}{l}{$\displaystyle 1-\left(1-\frac{3-2\nu+3\nu^2}{2(1+\nu)^2}\frac{1}{r^{*2}}\right)\frac{2}{r_1^*r_2^*}\cos(\theta_1-\theta_2)
    +\frac{1-\nu}{1+\nu}\frac{2r^{*2}(1-r^{*2})}{r_1^{*2}r_2^{*2}}\cos[2(\theta+\theta_1-\theta_2)]$}\\
    & \multicolumn{3}{l}{$\displaystyle -\frac{1-\nu}{1+\nu}\frac{4t^{*2}}{r_1^{*3}r_2^{*3}}
    \left[\cos(2\theta+3\theta_1-3\theta_2)
    +3r^{*2}\cos(4\theta+3\theta_1-3\theta_2)\right]$}\\
    & \multicolumn{3}{l}{$\displaystyle +\frac{2(3-\nu)}{(1+\nu)^2}\frac{1}{r^{*3}}\left[\cos\theta \log\frac{r_2^*}{r_1^*}+(\theta_1+\theta_2)\sin\theta\right]$}\\
    & \multicolumn{3}{l}{$\displaystyle + \frac{3(5-2\nu+\nu^2)}{(1+\nu)^2}\frac{1}{r^{*4}}\left[\cos(2\theta)\log(r_1^*r_2^*)-(\theta_1-\theta_2)\sin(2\theta) \right]$}\\
    \hline
    $\widetilde{\sigma}_{rr}^{(\mathrm{st})}$ 
    & \multicolumn{3}{l}{$\displaystyle 1 -\frac{2\cos\theta_1 \cos^2(\theta+\theta_1)}{r_1^*}
    - \frac{2\cos\theta_2 \cos^2(\theta-\theta_2)}{r_2^*}$}\\
    \hline
    $\widetilde{\sigma}_{rr,\mathrm{P}}^{(m)}$ 
    & $\begin{cases}
        \displaystyle -\frac{2f_{0,1}(r^* \omega^*)}{r^{*2}\omega^* g_{0,1}(\omega^*)} & (m=0)\\
        \displaystyle -\frac{4f_{m,1}(r^* \omega^*)f_{m,3}(\mu \omega^*)}{r^{*2}\omega^* d_m(\omega^*)} & (m\ge2)
    \end{cases}$
    & $\widetilde{\sigma}_{rr,\mathrm{S}}^{(m)}$ 
    & $\begin{cases}
        0 & (m=0)\\
        \displaystyle \frac{4f_{m,2}(\mu r^* \omega^*) f_{m,2}(\omega^*)}{r^{*2}\omega^* d_m(\omega^*)} & (m\ge2)
    \end{cases}$\\
    \hline\hline
    $\widetilde{\sigma}_{r\theta}^{(\mathrm{tr})}$ 
    & \multicolumn{3}{l}{$\displaystyle -\frac{3-2\nu+3\nu^2}{(1+\nu)^2}\frac{1}{r^{*2}r_1^*r_2^*}\sin(\theta_1-\theta_2)
    -\frac{1-\nu}{1+\nu}\frac{2r^{*2}(1-r^{*2})}{r_1^{*2}r_2^{*2}}\sin[2(\theta+\theta_1-\theta_2)]$}\\
    & \multicolumn{3}{l}{$\displaystyle +\frac{1-\nu}{1+\nu}\frac{4t^{*2}}{r_1^{*3}r_2^{*3}}\left[\sin(2\theta+3\theta_1-3\theta_2) +3r^{*2}\sin(4\theta+3\theta_1-3\theta_2)\right]$}\\
    & \multicolumn{3}{l}{$\displaystyle -\frac{2(3-\nu)}{(1+\nu)^2}\frac{1}{r^{*3}}\left[-\sin\theta\log\frac{r_2^*}{r_1^*} + (\theta_1+\theta_2)\cos\theta\right]$}\\
    & \multicolumn{3}{l}{$\displaystyle + \frac{3(5-2\nu+\nu^2)}{(1+\nu)^2}\frac{1}{r^{*4}}\left[\sin(2\theta)\log(r_1^*r_2^*)+(\theta_1-\theta_2)\cos(2\theta)\right]$}\\
    \hline
    $\widetilde{\sigma}_{r\theta}^{(\mathrm{st})}$ 
    & \multicolumn{3}{l}{$\displaystyle \frac{\cos\theta_1 \sin[2(\theta+\theta_1)]}{r_1^*}
    + \frac{\cos\theta_2 \sin[2(\theta-\theta_2)]}{r_2^*}$}\\
    \hline
    $\widetilde{\sigma}_{r\theta,\mathrm{P}}^{(m)}$
    & $\displaystyle - \frac{4f_{m,2}(r^* \omega^*)f_{m,3}(\mu \omega^*)}{r^{*2}\omega^* d_m(\omega^*)}$
    & $\displaystyle \widetilde{\sigma}_{r\theta,\mathrm{S}}^{(m)}$ 
    & $\displaystyle \frac{4f_{m,3}(\mu r^* \omega^*) f_{m,2}(\omega^*)}{r^{*2}\omega^* d_m(\omega^*)}$\\
    \hline
    $\widetilde{\sigma}_{\theta\theta}^{(\mathrm{tr})}$ 
    & \multicolumn{3}{l}{$\displaystyle 1
    -\left(1+\frac{3-2\nu+3\nu^2}{2(1+\nu)^2}\frac{1}{r^{*2}}\right)\frac{2}{r_1^*r_2^*}\cos(\theta_1-\theta_2)
    -\frac{1-\nu}{1+\nu}\frac{2r^{*2}(1-r^{*2})}{r_1^{*2}r_2^{*2}}\cos[2(\theta+\theta_1-\theta_2)]$}\\
    & \multicolumn{3}{l}{$\displaystyle +\frac{1-\nu}{1+\nu}\frac{4t^{*2}}{r_1^{*3}r_2^{*3}}
    \left[\cos(2\theta+3\theta_1-3\theta_2)
    +3r^{*2}\cos(4\theta+3\theta_1-3\theta_2)\right]$}\\
    & \multicolumn{3}{l}{$\displaystyle -\frac{2(3-\nu)}{(1+\nu)^2}\frac{1}{r^{*3}}\left[\cos\theta \log\frac{r_2^*}{r_1^*}+(\theta_1+\theta_2)\sin\theta\right]$}\\
    & \multicolumn{3}{l}{$\displaystyle - \frac{3(5-2\nu+\nu^2)}{(1+\nu)^2}\frac{1}{r^{*4}}\left[\cos(2\theta)\log(r_1^*r_2^*)-(\theta_1-\theta_2)\sin(2\theta) \right]$}\\
    \hline
    $\widetilde{\sigma}_{\theta\theta}^{(\mathrm{st})}$ 
    & \multicolumn{3}{l}{$\displaystyle 1 - \frac{2\cos\theta_1 \sin^2(\theta+\theta_1)}{r_1^*}
    - \frac{2\cos\theta_2 \sin^2(\theta-\theta_2)}{r_2^*}$}\\
    \hline
    $\widetilde{\sigma}_{\theta\theta,\mathrm{P}}^{(m)}$ 
    & $\begin{cases}
        \displaystyle -\frac{2f_{0,4}(r^* \omega^*)}{r^{*2}\omega^* g_{0,1}(\omega^*)} & (m=0)\\
        \displaystyle -\frac{4f_{m,4}(r^* \omega^*)f_{m,3}(\mu \omega^*)}{r^{*2}\omega^* d_m(\omega^*)} & (m\ge 2)
    \end{cases}$
    & $\widetilde{\sigma}_{\theta\theta,\mathrm{S}}^{(m)}$ 
    & $\begin{cases}
        0 & (m=0)\\
        \displaystyle -\frac{4f_{m,2}(\mu r^* \omega^*) f_{m,2}(\omega^*)}{r^{*2}\omega^* d_m(\omega^*)} & (m\ge 2)
    \end{cases}$\\
    \hline\hline
    \end{tabular}
\end{table}


\section{Results}\label{sec:results}
This section presents the results obtained from the theory in the previous section.
\begin{figure}[htbp]
    \centering
    \includegraphics[width=\linewidth]{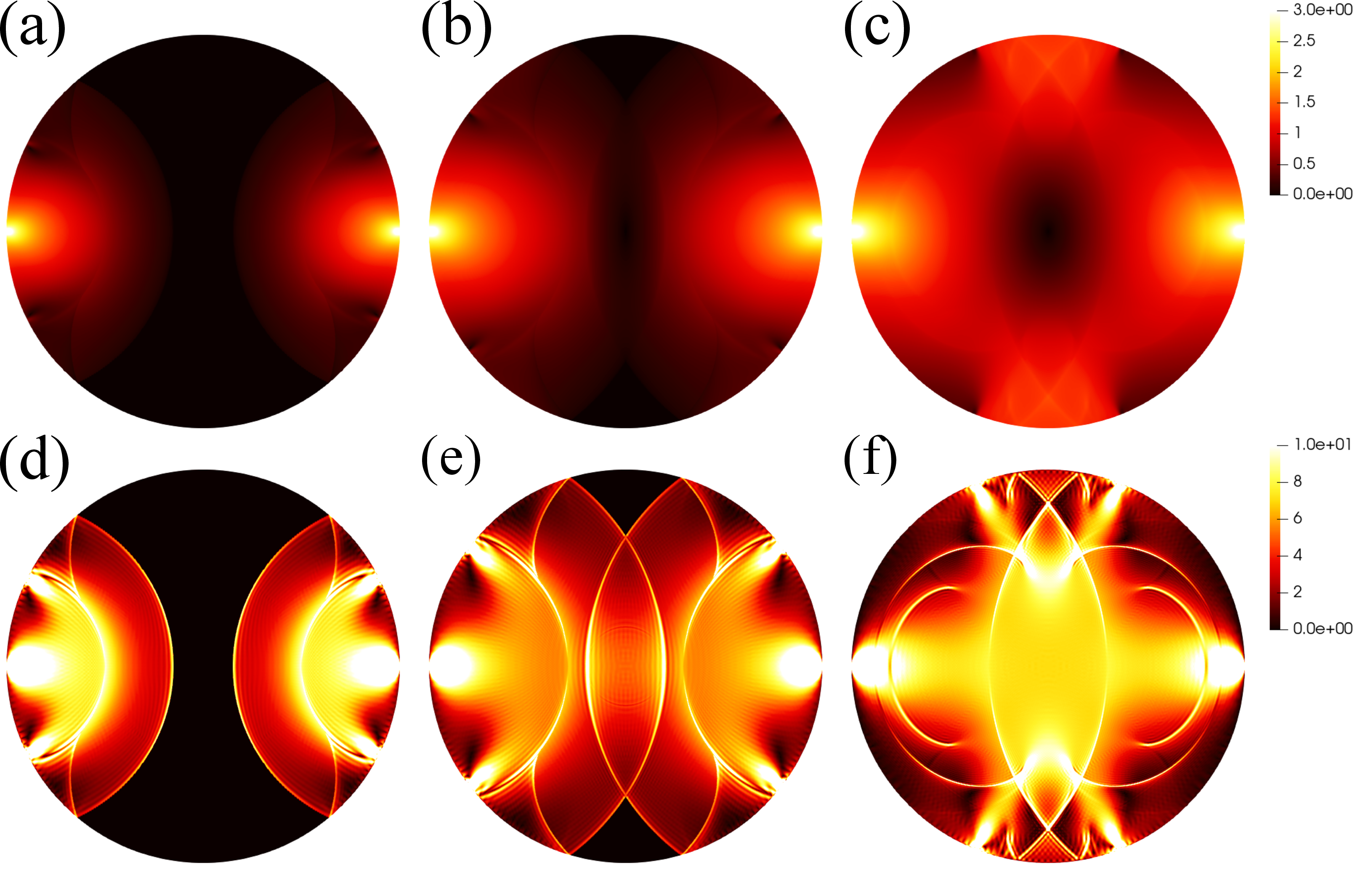}
    \caption{Density plots of (top) the magnitude of the (dimensionless) displacement $\widetilde{u}$ and (bottom) the (dimensionless) principal stress difference $\widetilde{\sigma}_{1}-\widetilde{\sigma}_{2}$ for $\nu=0.3$ at (a, d) $t^*=0.85$, (b, e) $1.2$, and (c, f) $2.2$.}
    \label{fig:evol}
\end{figure}
Figure \ref{fig:evol} shows the evolution of the magnitude of the strain $u$ and the principal stress difference $\sigma_{1}-\sigma_{2}$, calculating from Eq.~\eqref{eq:def_u_sigma}.
In the early stage of time evolution, the P- and S-waves propagate with speed $v_\mathrm{L}$ and $v_\mathrm{T}$, respectively, after the loading acts on the disk as shown in Fig.~\ref{fig:evol}~\cite{movie}.
When we focus on the principal stress difference (Fig.~\ref{fig:evol}(d)--(f)), the Rayleigh waves can also be observed on the surface, slightly delayed by the S-wave.
In addition, reflected waves (called the von Schmidt waves) appear from the intersections of the P-wave and the surface, which asymptotically approach the S-wave.
These behaviors are similar to those observed in the case of a semi-infinite space \cite{Fung}.
As time goes on, on the other hand, a lot of reflected waves appear because the system is finite.
When the waves reach the surface, they produce both P- and S-waves as reflected waves.
This means that the number of waves increases rapidly with time~\cite{movie}.
Then, the system converges to a static solution \eqref{eq:u_sigma_st} in the long-time limit.
In the following, we discuss each point.

The speed of the Rayleigh waves is understood as follows:
When we assume a flat surface, the propagation speed $c$ of the Rayleigh wave should satisfy \cite{Fung}
\begin{equation}
    \left(\frac{c}{c_\mathrm{T}}\right)^6 
    - 8\left(\frac{c}{c_\mathrm{T}}\right)^4 
    + 8(2+\nu)\left(\frac{c}{c_\mathrm{T}}\right)^2
    -8(1+\nu)=0.
    \label{eq:c_cT_eq}
\end{equation}
Solving numerically this equation, we get $c/c_\mathrm{T}\simeq 0.916$ for $\nu=0.3$.
Although this analysis is for the case of a flat system, the result is almost the same in this system.

\begin{figure}[htbp]
    \centering
    \includegraphics[width=0.8\linewidth]{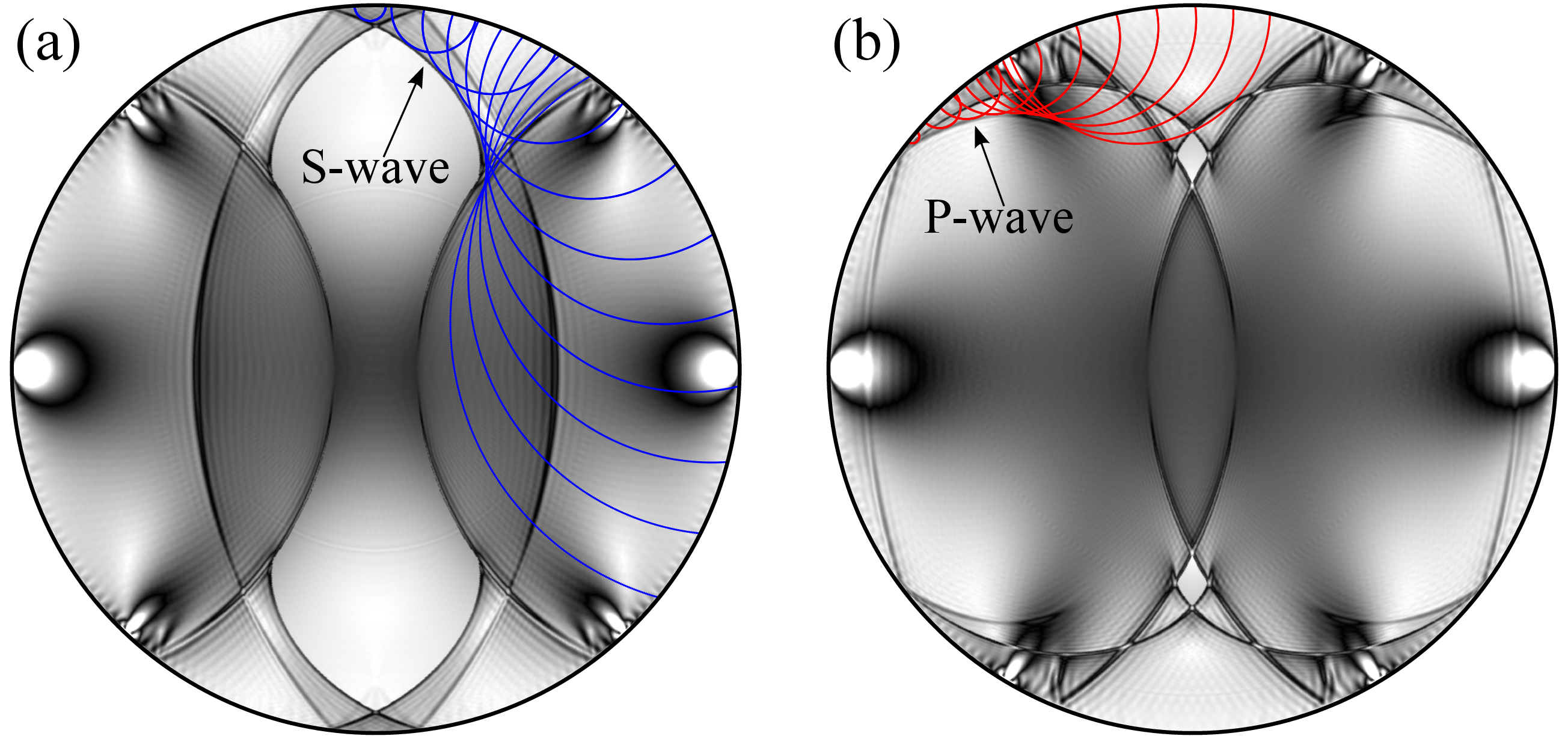}
    \caption{Density plots of the magnitude of the principal stress difference $\widetilde{\sigma}_{1}-\widetilde{\sigma}_{2}$ at (a) $t^*=1.50$ and (b) $1.90$, respectively.
    We also plot Eq.~\ref{eq:envelopes} for some values of $t^{\prime*}$.}
    \label{fig:envelopes}
\end{figure}

Next, let us consider the reflected waves from the intersections of the P-wave and the surface.
For simplicity, we only consider the wave propagating from $(r, \theta)=(a, 0)$ to the direction $\theta>0$.
At time $t^\prime$ ($0\le t^\prime < t$), the wavefront is at 
\begin{equation}
    \left(x_\mathrm{A}, y_\mathrm{A}\right)
    = \left(1-\frac{1}{2}\left(v_\mathrm{L}t^\prime\right)^2, 
    v_\mathrm{L}t^\prime \sqrt{1-\frac{1}{4}\left(v_\mathrm{L}t^\prime\right)^2}\right)a,
\end{equation}
in the dimensionless form.
From this point, the reflected P- and S-waves propagate with speed $v_\mathrm{L}$ and $v_\mathrm{T}$, respectively.
At the time $t$, the wavefronts of these reflected waves are, respectively, given by
\begin{align}
    (x - x_\mathrm{A})^2
    + (y - y_\mathrm{A})^2
    = 
    \begin{cases}
        v_\mathrm{L}^2(t-t^{\prime})^2 & (\mathrm{P-wave})\\
        v_\mathrm{T}^2(t-t^{\prime})^2 & (\mathrm{S-wave})
    \end{cases}
    .\label{eq:envelopes}
\end{align}
As the envelopes of these waves ($t^\prime \le t$), reflected waves can be observed as shown in Fig.~\ref{fig:envelopes}.
This is the origin of the von Schmidt waves.
It is noted that the reflected P-waves are rarely observed until the P-waves reach halfway past the disk.
This is because the envelope of the reflected P-wave is the P-wave itself.
After the halfway point, on the other hand, the reflected P-waves become observable due to geometrical reasons, as shown in Fig.~\ref{fig:envelopes}(b).
Of course, reflected waves are also generated by other waves travelling along the surface, e.g., S-waves.
However, the smaller the amplitude of the wave, the smaller the amplitude of the reflected wave.
The same discussion as the above treatment can be obtained, but this is not done here.

To understand the convergence of the waves, let us consider the tensile stress $\sigma_{yy}$.
This stress is known to be constant ($P_0/(\pi a)$) along the load axis under a static condition as discussed in Sect.~\ref{sec:static}.
\begin{figure}[htbp]
    \centering
    \includegraphics[width=0.5\linewidth]{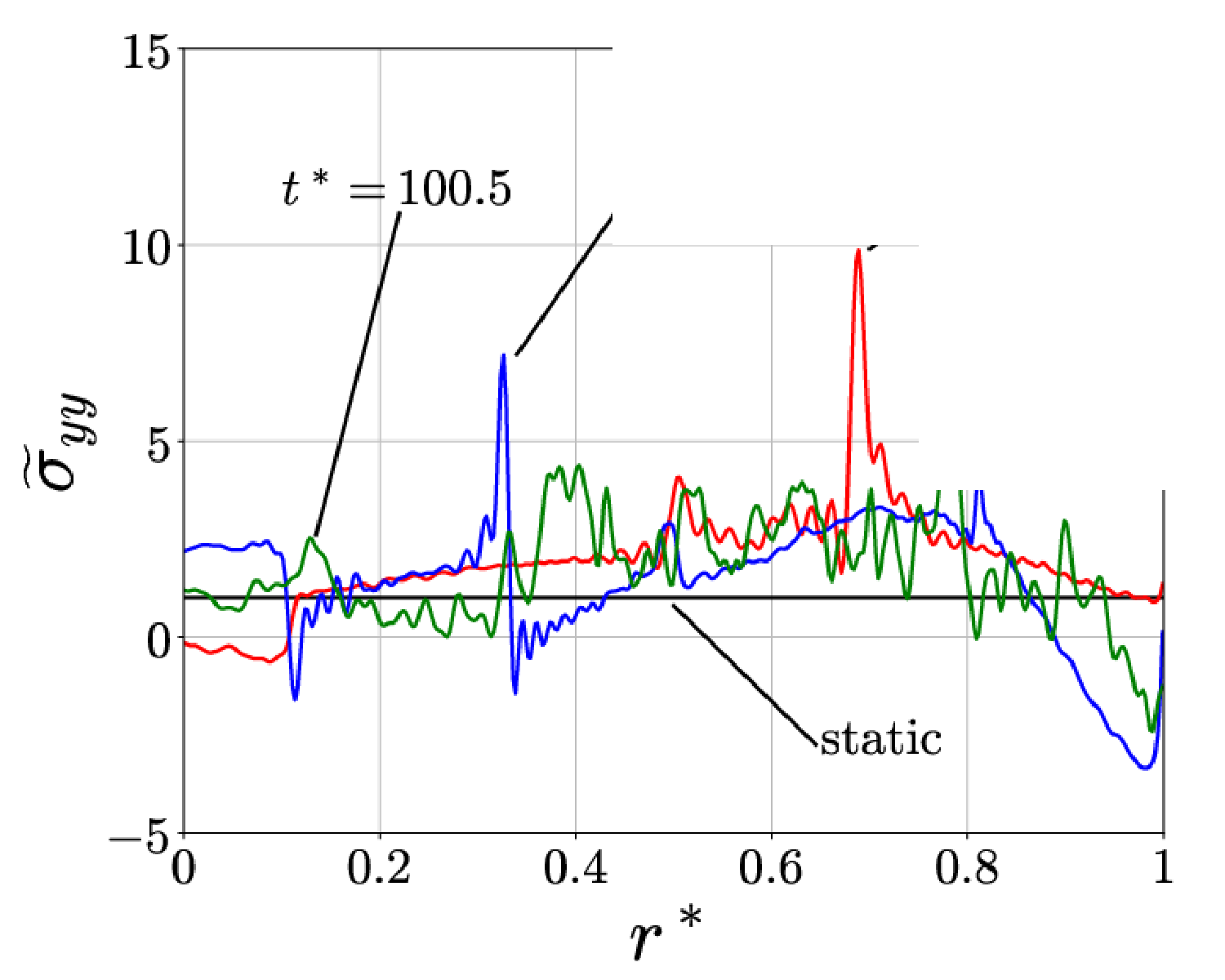}
    \caption{Plots of the dimensionless stress $\widetilde{\sigma}_{yy}$ for $\nu=0.3$ at $t^*=3.5$, $5.5$, and $100.5$.
    Here, the solid line represents the static solution $\widetilde{\sigma}_{yy}^{(\mathrm{st})}=1$.}
    \label{fig:evol_sigma_yy}
\end{figure}
Figure~\ref{fig:evol_sigma_yy} shows the time evolution of $\sigma_{yy}$ along the load axis ($0\le r\le a$, $\theta=0$).
In the early stage, the number of waves is small, therefore, the peaks of those waves stand out.
As time goes on, the number of reflected waves increases as discussed before.
Then, the magnitude of each peak becomes lower.
The waves are uniform in the long-time limit, where the number of waves becomes infinite.
This indicates, in other words, that tensile stress becomes many times larger than that predicted by the static solution before the long-time limit is satisfied. 
In this respect, the obtained analytical solutions quantitatively show that it is important to consider the safety factor in actual design.

\begin{figure}[htbp]
    \centering
    \includegraphics[width=0.8\linewidth]{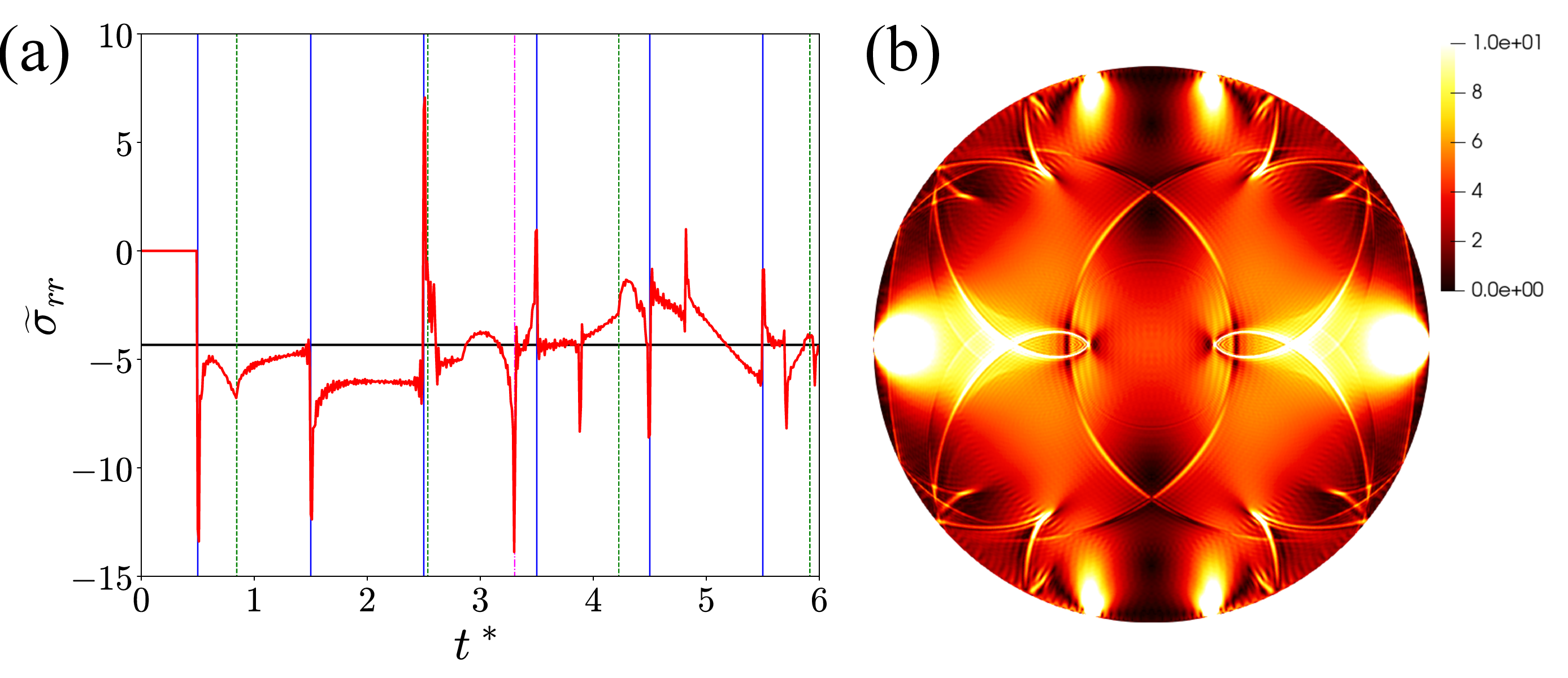}
    \caption{(a) Evolution of the (dimensionless) radial stress $\widetilde{\sigma}_{rr}$ at $(r^*,\theta)=(0.5, 0)$.
    The horizontal solid line represents the long-time limit obtained by Eqs.~\eqref{eq:u_sigma_m_lim}.
    The solid and dotted vertical lines mean the times when the P- and S-waves arrive, respectively.
    The dashed line is the time when S-waves generated by reflection of P-waves at $(r^*,\theta)=(1, \pm \pi/2)$ arrive at $(r^*,\theta)=(0.5, 0)$.
    (b) Density plot of the magnitude of the dimensionless principal stress difference at that time.}
    \label{fig:evolution_at_a_point}
\end{figure}

This behavior can be observed well when we fix a point.
Figure \ref{fig:evolution_at_a_point}(a) shows the evolution of the (dimensionless) radial stress at a point $(r^*,\theta)=(0.5, 0)$.
As shown in Fig.~\ref{fig:evolution_at_a_point}, the magnitude of the radial stress becomes large at every time P- or S-waves arrives at the point.
We can also observe that the stress converges to the long-time limit as shown in Fig.~\ref{fig:evolution_at_a_point}(a).
As time goes on, the peak at which a wave arrives becomes smaller. 
However, a large peak is occasionally observed, at which two waves arrive almost simultaneously (e.g., $t^*\simeq 2.5$).
Furthermore, not only direct waves, but also waves reflected and generated at $(r^*,\theta)=(1, \pm \pi/2)$ bring peaks. 
This is most noticeable in the peak at time $t^*\simeq 3.3$, which is the peak created by the S-waves produced by the reflection of the P-waves at $(r^*,\theta)=(1, \pm \pi/2)$.
This is clearly observed in Fig.~\ref{fig:evolution_at_a_point}(b).

\section{Conclusion and discussion}\label{sec:summary}
In this study, we undertook a thorough investigation into the dynamic response of a loaded disk, culminating in the derivation of analytical solutions. Our findings elucidate the initial propagation of P- and S-waves through the disk, concomitant with the discernible presence of Rayleigh waves at the surface. Notably, at the intersections of the P-wave and the surface, we observed the emergence of reflected waves, known as von Schmidt waves. Over time, owing to the finite nature of the system, the count of these reflected waves surges. We conducted a meticulous analysis of this phenomenon, revealing that reflected P-waves are a seldom occurrence until they traverse halfway across the disk. In the asymptotic long-time limit, the waves converge to a static solution.

Now, let us turn our attention to the convergence properties of the summations presented in Eq.~\eqref{eq:summation}. 
As deduced from Eq.~\eqref{eq:sigma_rr_m} in conjunction with Eq.~\eqref{eq:def_D_m}, it becomes apparent that the ratio of the numerator to the denominator in $\widetilde{\sigma}_{rr}^{(m)}$ approaches unity in close proximity to the disk's edge ($r^*\to 1$). This implies an imperative for substantially elevated values of $M\mathrm{max}$ and $N_\mathrm{max}$. It is worth noting that this condition extends to other parameters, given that the components of displacement and stress are intrinsically intertwined. This characteristic not only necessitates a substantial temporal investment but also introduces potential challenges in maintaining computational precision. From this vantage point, Finite Element Method (FEM) analysis may offer a synergistic approach to complement our analytical scrutiny.

Our work constitutes a revisitation of the seminal research conducted by Jingu \textit{et al.} \cite{Jingu85}. We have provided explicit and rigorously accurate expressions for solutions that were previously inadequately derived in their work. Additionally, we embarked upon a comprehensive discourse regarding the diverse waveforms encompassed within these solutions.

While we concentrated our efforts on scenarios involving point forces applied to a disk, it is crucial to underscore that our methodology transcends this specific configuration. By modulating the boundary conditions at the surface, our approach readily extends to systems subject to arbitrary boundary conditions. A similar adaptability holds true for the consideration of initial conditions, underscoring the versatility and applicability of our analytical framework.

\begin{acknowledgements}
The authors thank Hisao Hayakawa, Takashi Matsushima, Michio Otsuki, and Kenji Kurosaki for their helpful comments.
Numerical computation in this work was partially carried out at the Yukawa Institute Computer Facility. 
This research was partially supported by the Grant-in-Aid of MEXT for Scientific Research (Grant No.~JP20K14428 and No.~JP21H01006).
\end{acknowledgements}

\appendix
\section{Derivation of Eq.~\eqref{eq:u_sigma_st}}\label{sec:derivation}
In this Appendix, let us give the detailed derivations of the expressions of the static values \eqref{eq:u_sigma_st}.
Because we are only interested in the static values, we omit $\lim_{t\to\infty}$ in Eq.~\eqref{eq:u_sigma_m} in this Appendix.
Now, we focus on $\widetilde{\sigma}_{rr}^{(m)}$.
To calculate the summation $\sum_{m=2,4,\cdots}\widetilde{\sigma}_{rr}^{(m)}\cos(m\theta)$, it is convenient to use the following identity:
\begin{align}
    \sum_{m=2,4,\cdots}r^{*m}\cos(m\theta)
    &= \Re \sum_{m=2,4,\cdots}r^{*m}\mathrm{e}^{im\theta}
    = \Re \frac{1}{1-r^{*2}e^{2i\theta}}\nonumber\\
    &= \Re \frac{1}{r_1^*r_2^*}\mathrm{e}^{i(\theta_1-\theta_2)}
    = \frac{1}{r_1^*r_2^*}\cos(\theta_1-\theta_2),
    \label{eq:identity}
\end{align}
where $\Re$ represents the real part, and we have introduced $r_1$, $r_2$, $\theta_1$, and $\theta_2$ as shown in Fig.~\ref{fig:setup} in the second line of Eq.~\eqref{eq:identity}.
From this identity, the following is also obtained:
\begin{align}
    \sum_{m=2,4,\cdots}m r^{*m}\cos(m\theta)
    &= \Re \frac{2r^{*2}e^{2i\theta}}{\left(1-r^{*2}e^{2i\theta}\right)^2}
    = \Re \frac{2r^{*2}}{r_1^{*2}r_2^{*2}}\mathrm{e}^{2i(\theta + \theta_1-\theta_2)}\nonumber\\
    &= \frac{2r^{*2}}{r_1^{*2}r_2^{*2}}\cos[2(\theta + \theta_1-\theta_2)].
    \label{eq:identity2}
\end{align}
Using Eqs.~\eqref{eq:identity} and \eqref{eq:identity2}, one gets
\begin{equation}
    \sum_{m=2,4,\cdots}\widetilde{\sigma}_{rr}^{(m)}\cos(m\theta)
    = 1-\frac{\left(1-r^{*2}\right)^2\left[1-2r^{*2}-r^{*4}+2\cos(2\theta)\right]}{\left[1+r^{*4}-2r^{*2}\cos(2\theta)\right]^2}.
    \label{eq:sum_sigma_rr}
\end{equation}
Now, from the definition of $r^{*1}$, $r^{*2}$, $\theta_1$, and $\theta_2$ in Eqs.~\eqref{sec:def_r1_r2_theta1_theta2}, we can easily obtain
\begin{equation}
    \frac{\cos\theta_1 \cos^2(\theta+\theta_1)}{r_1^*}
    +\frac{\cos\theta_2 \cos^2(\theta-\theta_2)}{r_2^*}
    -\frac{1}{2}
    = \frac{\left(1-r^{*2}\right)^2\left[1-2r^{*2}-r^{*4}+2\cos(2\theta)\right]}{2\left[1+r^{*4}-2r^{*2}\cos(2\theta)\right]^2}.
    \label{eq:identity3}
\end{equation}
From Eqs.~\eqref{eq:sum_sigma_rr} and \eqref{eq:identity3} with Eq.~\eqref{eq:u_sigma_st_0}, we can reach Eq.~\eqref{eq:sigma_rr_st}.
After the similar calculations, we can also derive Eqs.~\eqref{eq:u_sigma_st}.

\section{Stress components in the Cartesian coordinates}\label{sec:Cartesian_polar}
In this Appendix, we present the expressions of the stress components in the Cartesian coordinates.
Once we obtain the expressions in the polar coordinates, it is straightforward to obtain those in the Cartesian coordinates as \cite{Timoshenko}
\begin{subequations}\label{eq:Cartesian_polar}
\begin{align}
    \widetilde{\sigma}_{xx}
    &= \frac{\widetilde{\sigma}_{rr}+\widetilde{\sigma}_{\theta\theta}}{2}
    +\frac{\widetilde{\sigma}_{rr}-\widetilde{\sigma}_{\theta\theta}}{2}\cos(2\theta) 
    - \widetilde{\sigma}_{r\theta}\sin(2\theta),\\
    \widetilde{\sigma}_{yy}
    &= \frac{\widetilde{\sigma}_{rr}+\widetilde{\sigma}_{\theta\theta}}{2}
    -\frac{\widetilde{\sigma}_{rr}-\widetilde{\sigma}_{\theta\theta}}{2}\cos(2\theta) 
    + \widetilde{\sigma}_{r\theta}\sin(2\theta),\\
    \widetilde{\sigma}_{xy}
    &= \widetilde{\sigma}_{r\theta}\cos(2\theta)
    + \frac{\widetilde{\sigma}_{rr}-\widetilde{\sigma}_{\theta\theta}}{2}\sin(2\theta).
\end{align}
\end{subequations}
Substituting Eqs.~\eqref{eq:u_sigma_st} into Eqs.~\eqref{eq:Cartesian_polar}, we can obtain
\begin{subequations}
\begin{align}
    \widetilde{\sigma}_{xx} 
    &= 1 -\left(\frac{2\cos^3\theta_1}{r_1} + \frac{2\cos^3\theta_2}{r_2}\right),\\
    \widetilde{\sigma}_{yy}
    &= 1 - \left(\frac{2\cos\theta_1 \sin^2\theta_1}{r_1} + \frac{2\cos\theta_2 \sin^2\theta_2}{r_2}\right),\\
    \widetilde{\sigma}_{xy}
    &= \frac{2\cos^2\theta_1 \sin\theta_1}{r_1} - \frac{2\cos^2\theta_2 \sin\theta_2}{r_2},
\end{align}
\end{subequations}
respectively \cite{Timoshenko}.
This shows that $\widetilde{\sigma}_{yy}$ becomes constant ($\widetilde{\sigma}_{yy}=1$) along the line parallel to the loading ($\theta_1=\theta_2=0$).

\section{Derivation of the speed of the Rayleigh wave}\label{sec:Rayleigh}
In this Appendix, we derive the speed of the Rayleigh wave.
The Rayleigh waves are produced by P- and S-waves, and propagate over the surface of the disk.
Let $c$ be the speed of the Rayleigh wave.
Now, it is convenient to have the origin on the surface of the disc, that is,
\begin{equation}
    r^\prime \equiv r-a,
\end{equation}
and use the coordinate system $(r^\prime, \theta)$.
Then, we assume that both the scalar and vector potentials are described by
\begin{subequations}
\begin{align}
    \phi &= \mathcal{A}_\mathrm{L}(r^\prime) \exp\left[i\omega \left(t-\frac{a\theta}{c}\right)\right],\label{eq:phi_Rayleigh}\\
    A &= \mathcal{A}_\mathrm{T}(r^\prime) \exp\left[i\omega \left(t-\frac{a\theta}{c}\right)\right],
\end{align}
\end{subequations}
respectively. 
Substituting Eq.~\eqref{eq:phi_Rayleigh} into the wave equation, the amplitude $\mathcal{A}_\mathrm{L}$ should satisfy
\begin{align}
    \frac{\partial^2 \mathcal{A}_\mathrm{L}(r^\prime)}{\partial r^{\prime2}} 
    +\frac{1}{a}\frac{\partial \mathcal{A}_\mathrm{L}(r^\prime)}{\partial r^\prime} 
    -\mathcal{A}_\mathrm{L}(r^\prime)\frac{\omega^2}{c^2 c_\mathrm{L}^2}(c_\mathrm{L}^2-c^2) = 0.
\end{align}
We should choose a solution, which does not diverge for $r\ll0$.
After this choice, the scalar potential is written as
\begin{equation}
    \phi
    = C_1 \exp\left[\frac{r^\prime}{2a}\left(\sqrt{1+4a^2\frac{\omega^2}{c^2 c_\mathrm{L}^2}(c_\mathrm{L}^2-c^2)}-1\right)\right]\exp(i\omega t)\exp\left(-\frac{i\omega a\theta}{c}\right).
    \label{eq:phi_Rayleigh}
\end{equation}
Similarly, the vector potential is given by
\begin{equation}
    A
    = C_2 \exp\left[\frac{r^\prime}{2a}\left(\sqrt{1+4a^2\frac{\omega^2}{c^2 c_\mathrm{T}^2}(c_\mathrm{T}^2-c^2)}-1\right)\right]\exp(i\omega t)\exp\left(-\frac{i\omega a\theta}{c}\right),
    \label{eq:A_Rayleigh}
\end{equation}
where $C_1$ and $C_2$ are constants.
If the second terms in the root of Eqs.~\eqref{eq:phi_Rayleigh} and \eqref{eq:A_Rayleigh} is much smaller than unity, we can rewrite them as
\begin{subequations}\label{eq:phi_A_Rayleigh}
\begin{align}
    \phi
    &= C_1 \exp\left[r^\prime a\left(\frac{\omega}{c}\right)^2\left(1-\frac{c^2}{c_\mathrm{L}^2}\right)\right]\exp(i\omega t)\exp\left(-\frac{i\omega a\theta}{c}\right),\\
    A
    &= C_2 \exp\left[r^\prime a\left(\frac{\omega}{c}\right)^2\left(1-\frac{c^2}{c_\mathrm{T}^2}\right)\right]\exp(i\omega t)\exp\left(-\frac{i\omega a\theta}{c}\right).
\end{align}
\end{subequations}
This solution needs to satisfy the boundary condition \eqref{eq:boundary_condition}.
After some calculations, the quantity $X\equiv c^2/c_\mathrm{T}^2$ should satisfy the following equation:
\begin{equation}
    X\left(X^3 - 8X^2 + 8(2+\nu)X-8(1+\nu) \right)=0.
\end{equation}
This is nothing but Eq.~\eqref{eq:c_cT_eq} in the main text.

\section*{Conflict of interest}
The authors declare that they have no conflict of interest.

\bibliographystyle{spphys}
\bibliography{references.bib}

\end{document}